\documentclass[12pt]{revtex4-2} 

\usepackage[utf8]{inputenc} 
\usepackage[vietnamese,english]{babel}

\usepackage{soul}
\usepackage{mathtools}
\usepackage{amsmath}  
\usepackage{amsfonts} 
\usepackage{graphicx} 
\usepackage{xcolor}
\usepackage{mathrsfs}

\begin{document}


\title{Vanishing in Fractal Space: \\ Thermal Melting and Hydrodynamic Collapse
}


\author{Trung V. Phan}
\email{tphan23@jhu.edu}
\affiliation{Department of Chemical and Biomolecular Engineering, \\ John Hopkins University, Baltimore, MD 21218, USA.}

\author{Truong H. Cai}
\affiliation{School of Engineering, Brown University, Providence, RI 02912, USA.}

\author{{\color{black}Van H. Do }}
\affiliation{{\color{black}Homer L. Dodge Department of Physics and Astronomy, University of Oklahoma, 440 W. Brooks St. Norman, OK 73019, USA.}}

\begin{abstract}
Fractals emerge everywhere in nature, exhibiting intricate geometric complexities through the self-organizing patterns that span across multiple scales. Here, we investigate beyond steady-states the interplay between this geometry and the vanishing dynamics, through phase-transitional thermal melting and hydrodynamic void collapse, within fractional continuous models. We present general analytical expressions for estimating vanishing times with their applicability contingent on the fractality of space. We apply our findings on the fractal environments crucial for plant growth: natural soils. We focus on the transport phenomenon of cavity shrinkage in incompressible fluid, conducting a numerical study beyond the inviscid limit. We reveal how a minimal collapsing time can emerge through a non-trivial coupling between the fluid viscosity and the surface fractal dimension.
\end{abstract}
 
\date{February 1st, 2024}

\maketitle 

\section{Introduction}

 {\color{black}Fractals often arise from the chaotic or stochastic self-organization of underlying dynamics \cite{rigon1994landscape,rodriguez1997fractal,baas2002chaos,bolliger2003self,ben2010role,isaeva2012self}, exhibiting self-similarity across different scales \cite{falconer2004fractal}.} {\color{black} Various} kinds of fractal spaces {\color{black} are found in nature \cite{bb1975objets,barnsley2014fractals}}, each distinguished by unique topological attributes \cite{weibel1991fractal,chen1998fractal}, in which the volume fractal dimension and the surface fractal dimension are the two most primary measures \cite{paumgartner1981resolution,friesen1987fractal,gimenez1997fractal}. Distinct topologies give rise to distinct physical manifestations \cite{phan2021curious,ao2023schrodinger}, and phenomena emerging within fractal spaces not only deviate significantly but also introduce new behaviors compared to their expected ones in conventional Euclidean spaces \cite{phan2020bacterial,martinez2022morphological,arellano2023habitat}. There is a huge potential for generalizations and discoveries of novel physical effects by studying well-established models under fractality \cite{tarasov2014flow,tarasov2014anisotropic,tarasov2015electromagnetic,tarasov2015elasticity,tarasov2015fractal,tarasov2016acoustic,naqvi2017electromagnetic}, following vector calculus on fractal space, a field that has seen significant development over the past two decades \cite{tarasov2011fractional,klafter2012fractional,tarasov2021general}. {\color{black} This} mathematical framework is founded on the assertion that mediums existing in fractal spaces can be effectively described using fractional continuous models \cite{tarasov2005continuous,tarasov2005possible}.

\begin{figure*}[!htbp]
\includegraphics[width=0.8\textwidth]{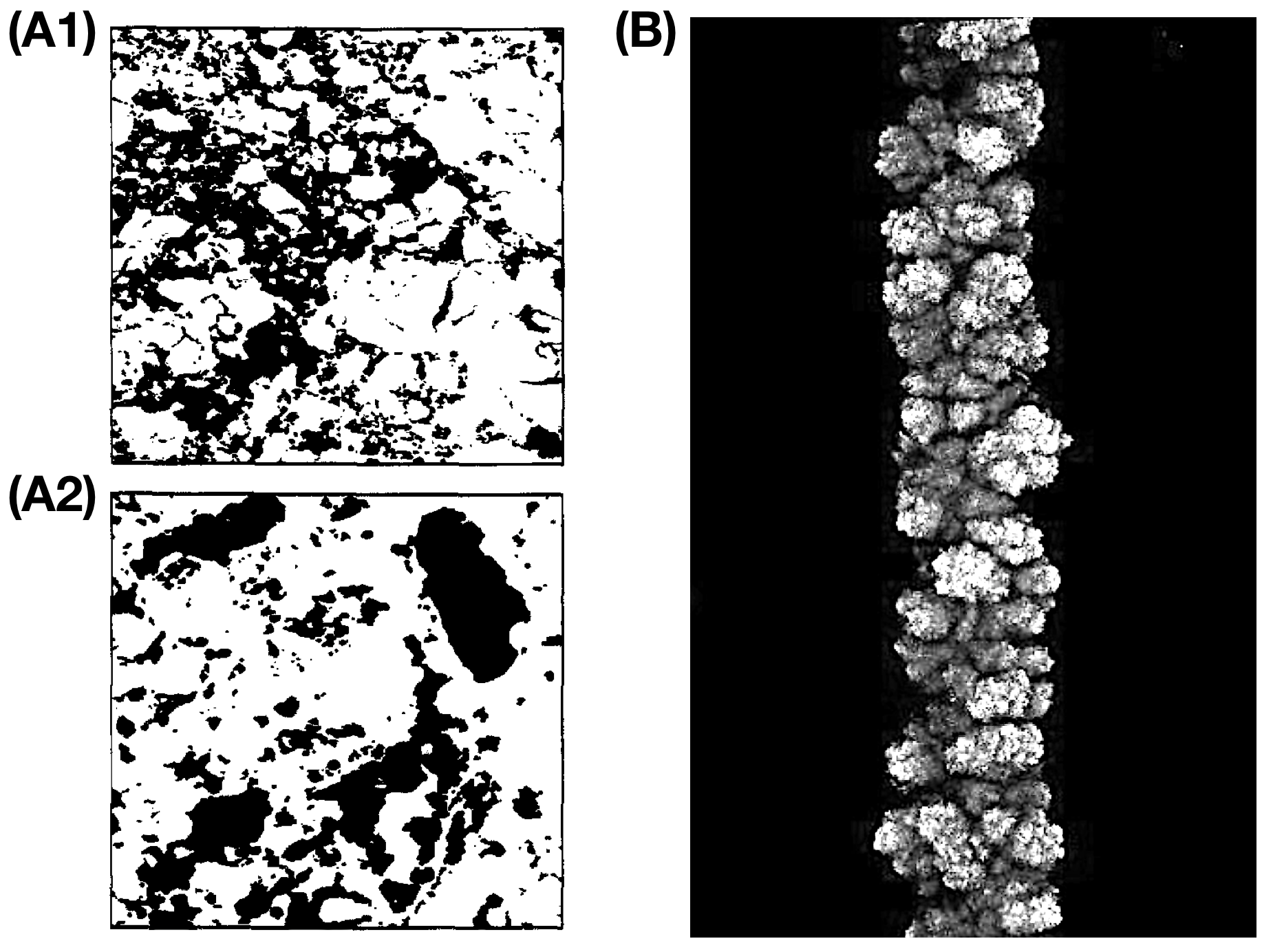}
\caption{\textbf{Fractal spaces and emergent fractional dynamics.} \textbf{(A)} Pores (black) distribution in natural soils, with volume fractal and surface fractal dimensionalities $\left(\mathscr{D}_\text{v},\mathscr{D}_\text{v}\right) \approx \left( 1.79,1.48\right)$ for \textbf{(A1)} and $\left( 1.69,1.24\right)$ for \textbf{(A2)}; images taken from \cite{gimenez1997fractal}. \textbf{(B)} Morphological instability and roughening of growing bacterial colonies (white) in transparent porous media; image taken from \cite{martinez2022morphological}.}
\label{fig00}
\end{figure*}

In Fig. \ref{fig00}A, we show a realization of fractal spaces found in the natural world, exemplified by the complex, multi-scaled, and multi-connected structures found in soils \cite{lai2016insight}. Soils play an irreplaceable role in facilitating plant growth. Not only do soils serve as a medium for plant anchorage, but their porous structures also allow for water retention and regulates its distribution, preventing excessive runoff and ensuring hydration for plant roots. Moreover, soils harbor a diverse microbiome that aids in nutrient cycling, disease suppression, and fostering symbiotic relationships, which are fundamental for plant health and resilience. Ultimately, the intricate combination of physical, chemical, and biological properties within soil makes them fundamental to sustaining plant life (and also the broader ecosystems depending on them).{\color{black} Let us briefly explain the ideas behind the conventional box-counting method used to measure fractality (technical details of how it can be done with images can be found in \cite{blacher1993use,chen1993calculation,huang1994can}). Fractal dimensions quantify how structures fill up space, describing how the distribution of details change with varying levels of magnification and spatial resolution \cite{grassberger1985generalizations}. First, we divide the fractal object into a grid of boxes of a fixed size. Then, we count the number of boxes that contain part of the soil pores. Repeat for different box sizes (spanning at least an order of magnitudes) and plot the relationship between box size and the number of boxes. The slope of the resulting log-log plot gives the volume fractal dimension $\mathscr{D}_\text{v}$. For the area fractal dimension $\mathscr{D}_\text{s}$, we replace boxes with squares. For natural soils, we can obtain small uncertainties with this simple method:} $\mathscr{D}_\text{v}$ ranges from $1.69 {\color{black} \pm 0.06}$ to $1.79 {\color{black} \pm 0.06}$, while $\mathscr{D}_\text{s}$ spans from $1.24 {\color{black} \pm 0.01}$ to $1.48 {\color{black} \pm 0.01}$ \cite{gimenez1997fractal}.

Fig. \ref{fig00}B showcases an interesting emergent phenomenon: the growth of microbes within a fractal environment \cite{martinez2022morphological}. This observation has become possible only recently, thanks to the development of three-dimensional printing techniques that allow for the creation of a self-healing ``transparent soil'' \cite{bhattacharjee2019bacterial,bhattacharjee2021chemotactic}. In there, chemotactic bacteria populations interact dynamically with their geometric surroundings \cite{lambert2010collective,morris2017bacterial}, aggregating into colonies that ``resemble broccoli in shape'', featuring irregular, self-similar surfaces \cite{park2003influence,phan2020bacterial}. Undoubtedly, numerous fascinating physical phenomena within fractal spaces remain unexplored, awaiting discoveries! 

Here, we explore the interaction between fractal geometries and the important physical phenomenon of vanishing dynamics, observed in phase-transitional thermal melting \cite{mccue2008classical} and hydrodynamic void collapse \cite{walton1984sonoluminescence,margulis1995sonochemistry,young1989cavitation,barber1991observation}. 
This area has received relatively much less attention compared to the study of growing dynamics \cite{vicsek1992fractal}. Thermal melting and void collapse can be viewed as simplifications of heat and fluid permeation within a fractal structure \cite{sreenivasan1994fractals}. We use nondimensionalization \cite{manteca2012mathematical} to acquire scale-free interpretations, rendering it more natural to facilitate comparisons of results acros
s various fractal dimensionalities. We begin with the classical problem of determining the time it takes for a spherical solid at its melting temperature, enclosed by warmer media, to completely {\color{black} liquify} \cite{vuik1993some}. Even without an exact solution, we can still approximate the process as quasi-stationary to estimate the melting time. We assess the validity of this approximation and provide a general analytical estimation for the melting time in fractal spaces. Then, we delve into the hydrodynamics of vacuum cavity contraction within an incompressible fluid. We conduct analytical and numerical investigations of collapsing time in fractal spaces, extending beyond the inviscid regime \cite{marshall2001inviscid}. We discover a curious optimization of mass transport in relative units that arises from the non-trivial coupling between fluid viscosity and the surface fractal dimension, manifesting only when both of these physical factors contribute. Our finding highlights the potential of using fractality as a critical engineering parameter in hydrodynamical applications. Finally, to relate these mathematical results to real-world scenarios, we probe the behavior of vanishing dynamics within a range of fractal dimensionalities similar to those observed in natural soils, i.e. with $\mathscr{D}_\text{v}\in [1.69,1.79]$ and $\mathscr{D}_\text{s} \in [1.24,1.48]$ \cite{gimenez1997fractal}. We emphasize that the fundamental assumption of this work -- fractional continuous models can be a reasonable description of mediums on fractal spaces \cite{tarasov2005continuous,tarasov2005possible} -- should not be view as definitive, but rather as a heuristic tool offering us some glimpses into the complexity that arises in physics within fractal spaces.

\section{The Basics of Vector Calculus in Fractal Spaces}

Fractal spaces introduce novel physics that demand a corresponding adaptation of mathematics. Specifically, the standard vector calculus used in conventional Euclidean spaces must be appropriately modified to accurately describe these unique environments. Let us start with some geometric characteristics of fractal spaces. Consider a spherical ball $\mathcal{S}$ of radius $r$ in this space, with surface $\partial \mathcal{S}$. The fractal dimensionalities $\text{dim}\left( \mathcal{S}\right) = \mathscr{D}_{\text{v}}$ and $\text{dim}\left( \partial \mathcal{S}\right) = \mathscr{D}_{\text{s}}$ can be non-integers. The fractality associated with the radial direction is $\alpha_r = \mathscr{D}_{\text{v}} - \mathscr{D}_{\text{s}}$. In general, the volume $\mathcal{V}(r)$ and the surface area of this ball $\mathcal{A}(r)$ follow power-law dependencies, i.e. $\mathcal{V}(r) \propto r^{\mathscr{D}_\text{v}}$ and $\mathcal{A}(r) \propto r^{\mathscr{D}_\text{s}}$, with the coefficients of proportionality depend on the particular spatial topology. For simplicity, we just assume these functions to be analytical continuations of those defined for hyperspheres \cite{das1990hyperspheres}:
\begin{equation}
\mathcal{V}(r) = \frac{\displaystyle \pi^{\mathscr{D}_{\text{v}}/2}}{\displaystyle \Gamma \left( \frac{\mathscr{D}_{\text{v}}+2}{2}\right)} r^{\mathscr{D}_{\text{v}}}  \ \ , \ \ \mathcal{A}(r) = \frac{\displaystyle 2\pi^{(\mathscr{D}_{\text{s}}+1)/2}}{\displaystyle \Gamma \left( \frac{\mathscr{D}_{\text{s}}+1}{2}\right)} r^{\mathscr{D}_{\text{s}}} \ \ .
\label{fractal_volume_surface}
\end{equation}
{\color{black} Here $\Gamma(...)$ refers to the Gamma function, an extension of the factorial function to complex numbers \cite{edwards2001riemann}.} The constraint of embedding in the physical three-dimensional Euclidean space bounds the fractal dimensions, i.e. $\mathscr{D}_{\text{v}}\leq 3$, $\mathscr{D}_{\text{s}} \leq 2$, and $\mathscr{D}_{\text{s}} \leq \mathscr{D}_{\text{v}}$. {\color{black} As a demonstration}, the fractality of natural soils satisfy these \cite{gimenez1997fractal}.
The dimensionless prefactors in Eq. \eqref{fractal_volume_surface} exhibit monotonic increase within these ranges. 

In this work, we only study spherical symmetric systems. The $\vec{\nabla}$ operator in this fractal space acting on radial scalar function $S(r)$ gives:
\begin{equation}
\vec{\nabla} S(r) = \frac{\displaystyle \Gamma\left(\frac{\alpha_r}{2}\right)}{\displaystyle \pi^{\alpha_r/2}} \frac1{\displaystyle r^{\alpha_r-1}} \partial_r S(r) \hat{r} \ \ ,
\label{fractal_grad}
\end{equation}
and on radial vector function $\vec{V}(r) = V(r) \hat{r}$ gives:
\begin{equation}
\vec{\nabla} . \vec{V}(r) = \frac{\displaystyle \Gamma\left( \frac{\mathscr{D}_{\text{v}}}{2}\right)}{\displaystyle \pi^{(\alpha_r - 1)/2} \Gamma\left( \frac{\mathscr{D}_{\text{v}} - \alpha_r +1}{2}\right)} \left( \frac1{\displaystyle 
 r^{\alpha_r - 1}} \partial_r + \frac{\displaystyle \mathscr{D}_{\text{v}} - \alpha_r}{\displaystyle 
 r^{\alpha_r}} \right) V(r) \ \ .
\label{fractal_div}
\end{equation}
For the tensor $\vec{\nabla}\vec{V}$, we treat each component of as a scalar and apply Eq. \eqref{fractal_grad}. We show how the Laplacian $\Delta = \vec{\nabla}.\vec{\nabla}$ acts on these in Appendix \ref{Lap}. These results are reported in \cite{tarasov2014flow}, relying on an extension of the Green–Gauss theorem to fractal objects through fractional integrals in Euclidean spaces \cite{ostoja2009continuum}. 

From a theoretical point of view, this novel framework of vector calculus holds promise for unveiling previously inaccessible insights into the behavior and properties of complex geometric structures, enabling a more comprehensive understanding of natural phenomena governed by fractal geometries. Moreover, it opens up avenues for the development of specialized mathematical tools tailored to address the unique challenges posed by these intricate spatial arrangements. This approach not only enriches the theoretical landscape but also has the potential to yield practical applications in various scientific and engineering disciplines where fractal geometries play a crucial role.

\section{Vanishing Phenomenon in Euclidean Spaces}

Before embarking on an exploration in fractal spaces, it is important to revisit the  vanishing dynamics of thermal melting and hydrodynamic collapse occurring in the physical three-dimensional Euclidean space, characterized by $\left( \mathscr{D}_{\text{v}}, \mathscr{D}_{\text{s}} \right) = \left( 3,2 \right)$.

\subsection{Thermal Melting \label{sec_therm_melt}}

Imagine a solid sphere initially of radius $R(t)\Big|_{t=0}$ at its melting temperature $T_{\text{m}}$, having a latent heat of $L$ for the solid-to-liquid phase-transition. This sphere is placed deep within a warmer liquid medium of heat capacity $C_{\text{l}}$ and conductivity $\kappa$, with an ambient temperature $T_{\infty} > T_{\text{m}}$. We consider this scenario in the standard three-dimensional open world we are familiar with.  In the thermodynamics literature, this setup is referred to as the classical two-phase Stefan problem with a moving boundary \cite{vuik1993some}. While it is really hard to find exact solutions for this problem, we can still make approximate calculations to get a reasonable estimate for how long it takes for the sphere to melt \cite{vpho2010,mccue2008classical}.

We choose a spherical coordinate system which origin coincides with the center of the melting solid sphere, and impose rotational symmetry, meaning all dynamical spatio-temporal functions vary in space only by their radial position. To nondimensionalize this heat transport phenomenon, we define the dimensionless temperature $\tilde{T}$ from physical temperature $T$, the dimensionless radius $\tilde{r}$ from physical radius $r$, the dimensionless time $\tilde{t}$ from physical time $t$, and the dimensionless heat $\tilde{Q}$ from physical heat $Q$:
\begin{equation}
\tilde{T} \equiv \frac{T- T_\text{m}}{T_\infty - T_\text{m}} \ \ , \ \ \tilde{r} \equiv \frac{r}{R(0)} \ \ , \ \ \tilde{t} \equiv t\left[ \frac{LR^2(0)}{\kappa \left( T_{\infty} - T_{\text{m}}\right)}\right]^{-1} \ \ , \ \ \frac{d}{d\tilde{t}} \tilde{Q} \equiv \frac1L \frac{d}{dt} Q \ \ .
\end{equation}
For examples, the dimensionless radius $\tilde{R}(\tilde{t})$ of the melting sphere should be defined by the ratio $R(t)/R(0)$, and we can set $\tilde{\kappa}=1$ and $\tilde{L}=1$. From now on, the tilde-hat notation $\tilde{ \bullet }$ will always refer to the dimensionless versions of physical quantities $\bullet$.

Let us now show how thermal melting can be approximated. For the surrounding liquid having very small heat capacity $C_{\text{l}} \rightarrow 0$, we can assume quasi-stationary temperature $\partial_{\tilde{t}} \tilde{T} \propto \tilde{\Delta} \tilde{T} \rightarrow 0$, thus the total incoming heat flux $\tilde{\vec{J}} = \mathcal{A}(\tilde{r})  \left( \tilde{\kappa} \tilde{\vec{\nabla}} \tilde{T} \right)$ should be a radial constant and can be estimated:
\begin{equation}
\tilde{J}(\tilde{t}) = - \tilde{T}(\tilde{r}',\tilde{t})\Big|^{\tilde{r}'=\infty}_{\tilde{r}'=\tilde{R}(\tilde{t})}\left[ \int^{\infty}_{\tilde{R}(\tilde{t})} d\tilde{r}' \mathcal{A}^{-1}(\tilde{r}') \right]^{-1} = - \left[ \int^{\infty}_{\tilde{R}(\tilde{t})} d\tilde{r}' \mathcal{A}^{-1}(\tilde{r}') \right]^{-1} \ \ .
\label{heat_flux_constant_main}
\end{equation}
Here $\mathcal{A}(\tilde{r})=4\pi \tilde{r}^2$ is the surface area of a spherical boundary with radius $\tilde{r}$, following from Eq. \eqref{fractal_volume_surface}. We do the integration and obtain $\tilde{J}(\tilde{t}) = -4\pi \tilde{R}(\tilde{t})$. The negative sign indicates this heat rate is going toward the solid sphere. This amount of heat rate absorbed by the sphere makes it melt, which allows us to obtain a differential equation describing how its size change with time:
\begin{equation}
\frac{d}{d\tilde{t}} \mathcal{V}(\tilde{r}) \Big|_{\tilde{r} = \tilde{R}(\tilde{t})} = \tilde{J}(\tilde{t}) \ \ \Longrightarrow \ \ \frac{d}{d\tilde{t}} \tilde{R}(\tilde{t}) = -\frac1{\tilde{R}(\tilde{t})} \ \ .
\end{equation}
We have used Eq. \eqref{fractal_volume_surface} to get the volume of a spherical region with radius $\tilde{r}$, $\mathcal{V}(\tilde{r})=4\pi \tilde{r}^3/3$. The total melting time $\tilde{\tau}_{\text{m}}$ can be found by taking an integration from $\tilde{R}(0)=1$ to $\tilde{R}(\tilde{\tau}_{\text{m}})=0$:
\begin{equation}
\tilde{\tau}_{\text{m}} = -\int^0_1 d\tilde{R}' \frac1{\tilde{R}'(\tilde{t})} = \frac12 \ \ .
\label{melt_time_3D}
\end{equation}
For more details, we show this derivation of the total melting time in three-dimensional Euclidean space, without nondimensionalization, in Appendix \ref{therm_melt_3D}.

In two-dimensional Euclidean space, the constant radius boundary is a circle $\mathcal{A}(\tilde{r}) \propto \tilde{r}$, hence the integral $\int d\tilde{r}' ...$ in Eq. \eqref{heat_flux_constant_main} does not converge. It means our quasi-stationary assumption fails, since as we approach that limit the total heat transfer from the environment to the solid sphere goes to zero. However, this is just a failure in the quantitative estimation, but not in the qualitative interpretation -- physically, the melting time becomes extremely long. 

\subsection{Hydrodynamic Collapse \label{sec_hyd_collapse}}

Under high-frequency perturbation e.g. ultrasonication, sonochemistry, lithotripsy, laser-surgery \cite{walton1984sonoluminescence,margulis1995sonochemistry,young1989cavitation,barber1991observation}, a void can emerge inside a liquid body. We are interested in the temporal dynamic of such a cavity. The hydrodynamic invasion into vacant spaces could also occur under analogous conditions.

Consider a vacuum bubble initially of radius $R(t)\Big|_{t=0}$ surrounded by an incompressible fluid with mass density $\rho$ and kinematic shear-viscosity $\mu$. The ambient pressure applied on the system, as measured from far-away infinity, is $P_{\infty}$. We assume that, in the beginning, the whole system can be at rest $dR(t)/dt\Big|_{t=0}=0$. The scenario we describe here happens in the standard three-dimensional open world we experience daily. For an incompressible liquid medium at the inviscid regime $\mu \rightarrow 0$, this is known as the Rayleigh problem \cite{rayleigh1917viii}. 

Let us define a characteristic velocity $U$ and a Reynolds number $\mathscr{R}$ as follows:
\begin{equation}
U \equiv \left(\frac{P_{\infty}}{\rho}\right)^{1/2} \ \ , \ \ \mathscr{R} \equiv \frac{UR(0)}{\mu} \ \ .
\label{U_Re}
\end{equation}
We choose a spherical coordinate system {\color{black} whose} origin coincides with the center of
the void, and impose rotational symmetry, meaning all dynamical spatio-temporal functions vary in space only by their radial position.
We obtain the nondimensionalization of the hydrodynamic Navier-Stokes equation \cite{salvi1998navier}:
\begin{equation}
\partial_{\tilde{t}} \tilde{\vec{v}} + \left( \tilde{\vec{v}} . \tilde{\vec{\nabla}} \right) \tilde{\vec{v}} = - \tilde{\vec{\nabla}} \tilde{P} + \tilde{\vec{\nabla}} .\tilde{\sigma} \ \ .
\label{NS_eq_dimless}
\end{equation}
where $P(r,t)$ is the pressure and $\sigma(r,t)$ is the deviatoric stress-tensor \cite{stokes2007theories}:
\begin{equation}
\tilde{\sigma} \propto  \vec{\nabla}\vec{v} + (\vec{\nabla}\vec{v} )^{T} \ \ \Longrightarrow \ \ \tilde{\vec{\nabla}}.\tilde{\sigma} = \frac1{\mathscr{R}} \tilde{\Delta} \tilde{\vec{v}} \ \ .
\label{deviatoric_dimless}
\end{equation}
Here we introduce dimensionless velocity vector field $\tilde{\vec{v}}$, radius $\tilde{r}$, time $\tilde{t}$ and pressure $\tilde{P}$:
\begin{equation}
\tilde{\vec{v}} \equiv \frac{\vec{v}}{U} \ \ , \ \ \tilde{r} \equiv \frac{r}{R(0)} \ \ , \ \ \tilde{t} \equiv t \left[ \frac{R(0)}{U} \right]^{-1} \ \ , \ \ \tilde{P} \equiv \frac{P}{\rho U^2} = \frac{P}{P_{\infty}} \ \ .
\label{dimless_collapse}
\end{equation}
For examples, the dimensionless radius $\tilde{R}(\tilde{t})$ of the collapsing void should be defined by the ratio $R(t)/R(0)$, and we can set $\tilde{\rho}=1$. 

From the incompresibility condition $\tilde{\vec{\nabla}} . \tilde{\vec{v}} = 0$ , we can {\color{black} express the radial velocity field $\tilde{v}(\tilde{r}, \tilde{t})$ everywhere in terms of how the cavity size $\tilde{R}(\tilde{t})$ changes with time}:
\begin{equation}
\tilde{v}(\tilde{r},\tilde{t}) = \left[ \frac{\tilde{R}(\tilde{t})}{\tilde{r}} \right]^2 \frac{d}{d\tilde{t}} \tilde{R}(\tilde{t}) \ \ . 
\label{vec_field_dimless}
\end{equation}
At the surface of the spherical bubble $\tilde{r}=\tilde{R}(\tilde{t})$, the pressure is equal to the viscous normal-stress from Eq. \eqref{deviatoric_dimless}:
\begin{equation}
\tilde{P}(\tilde{r})\Big|_{\tilde{r}=\tilde{R}(\tilde{t})} = \tilde{\sigma}_{\tilde{r} \tilde{r}}(\tilde{r},\tilde{t}) \Big|_{\tilde{r}=\tilde{R}(\tilde{t})} = - \frac{4}{\mathscr{R}} \frac1{\tilde{R}(\tilde{t})} \frac{d}{d\tilde{t}} \tilde{R}(\tilde{t}) \ \ .
\end{equation}
{\color{black} This is the dynamic equilibrium at the interface, where the normal forces must be balanced as the interface itself has no mass \cite{brennen2005fundamentals}.} We apply the fluid velocity field as found in Eq. \eqref{vec_field_dimless} to the nondimensionalized Navier-Stokes equation Eq. \eqref{NS_eq_dimless}, then radially integrate both side from the bubble surface $\tilde{r} = \tilde{R}(\tilde{t})$ to far-away infinity $\tilde{r} \rightarrow \infty$ to obtain:
\begin{equation}
\tilde{R}(\tilde{t}) \frac{d^2}{d\tilde{t}^2} \tilde{R}(\tilde{t}) + \frac32 \left[\frac{d}{d\tilde{t}} \tilde{R}(\tilde{t}) \right]^2 + \frac{4}{\mathscr{R}} \frac1{\tilde{R}(\tilde{t})}  \frac{d}{d\tilde{t}} \tilde{R}(\tilde{t}) + 1 = 0 \ \ .
\label{collapse_ode_main}
\end{equation}
This is known as the Rayleigh-Plesset equation \cite{plesset1949dynamics}. We proceed with the inviscid limit $\mathscr{R}\rightarrow \infty$. From the initial conditions $\tilde{R}(0)=1$ and  $d/d\tilde{t} \tilde{R}(\tilde{t})\Big|_{\tilde{t}=0} = 0$ and after a temporal integration, we arrive at:
\begin{equation}
 \frac{d}{d\tilde{t}} \tilde{R}(\tilde{t}) = -\left( \frac{2}{3} \right)^{1/2} \left\{ \left[ \frac{1}{\tilde{R}(\tilde{t})} \right]^3 - 1 \right\}^{1/2} \ \ .
\end{equation}
The void collapsing time $\tilde{\tau}_\text{c}$ can therefore be found to be:
\begin{equation}
\tilde{\tau}_{\text{c}} = - \left( \frac{3}{2} \right)^{1/2}\int_1^0 d\tilde{R}' \left\{ \left[ \frac{1}{\tilde{R}'} \right]^3 - 1 \right\}^{-1/2} 
= \frac{\displaystyle \left(\frac{3\pi}{2}\right)^{1/2} \Gamma \left( \frac56 \right)}{\displaystyle 3 \Gamma \left(\frac43 \right)} \ \ .
\label{collapsing_time_3D_main}
\end{equation}
For more details, we show this derivation of the vanishing time of vacuum cavity in three-dimensional Euclidean space, without nondimensionalization, in Appendix \ref{hyd_collapse_3D}. For the situation in viscous fluid, see \cite{bogoyavlenskiy1999differential}.

It can be shown that, in two-dimensional Euclidean space, Eq. \eqref{collapse_ode_main} becomes.
\begin{equation}
- \frac12 \left[ \frac{d}{d\tilde{t}} \tilde{R}(\tilde{t})\right]^2 + \frac{2}{\mathscr{R}} \frac1{\tilde{R}(\tilde{t})} \frac{d}{d\tilde{t}} \tilde{R}(\tilde{t}) + 1 = 0 \ \ .
\label{plesset_rayleigh_2D}
\end{equation}
We refer to Section \ref{hyd_frac_main} and Appendix \ref{hyd_fractal_supp} for the derivation of this result. {\color{black} Notably, the assumption for the fluid being initially at rest is inconsistent with this hydrodynamic model.} Either we need to change the initial condition or we need to include other effect e.g. the compressibility of real fluid. {\color{black} In this paper, we will not delve deeper -- we only mention it as a potential direction for future exploration.}

{\color{black} In recent developments, Zhang \cite{zhang2023unified} has attempted to synthesize insights from various models for hydrodynamic voids. This new framework accommodates a range of factors including different boundary types, bubble interactions, background flow, gravitational effects, migration, fluid compressibility, viscosity, and surface tension. To our knowledge, however, none of these models have yet to consider the interaction between complex fractal geometry and the collapse of hydrodynamic voids. The significance of fractal geometry within fluid mechanics is typically discussed in relation to permeability models \cite{huang2023vuggy} and drainage dynamics \cite{zhao2023drainage} for vuggy porous media. A comprehensive understanding of this interaction would yield substantial benefits for engineering disciplines where the transport in porous media plays a critical role, e.g. oil extraction \cite{shang2022thermal,sirine2022numerical,nandlal2019drained}.}

\section{Thermal Melting in Fractal Spaces}

We now extend our consideration to fractal spaces characterized by general non-integer dimensionalities $\left( \mathscr{D}_\text{v},\mathscr{D}_\text{s} \right) \in \left([0,3],[0,2]\right)$ with $\mathscr{D}_\text{v} \geq \mathscr{D}_\text{s}$. We proceed with calculations similar to those outlined in Section \ref{sec_therm_melt}. The primary difference lies in the utilization of fractional gradient and divergence operators as described in Eq. \eqref{fractal_grad} and Eq. \eqref{fractal_div}. Same as Eq. \eqref{U_Re} and Eq. \eqref{dimless_collapse},
we define:
\begin{equation}
U \equiv \left(\frac{P_{\infty}}{\rho}\right)^{1/2} \ \ , \ \ \mathscr{R} \equiv \frac{UR^{2\alpha_r-1}(0)}{\mu} \ \ ,
\end{equation}
and:
\begin{equation}
\tilde{\vec{v}} \equiv \frac{\vec{v}}{U} \ \ , \ \ \tilde{r} \equiv \frac{r}{R(0)} \ \ , \ \ \tilde{t} \equiv t \left[ \frac{R^{\alpha_r}(0)}{U} \right]^{-1}  \ \ , \ \ \tilde{P} \equiv \frac{P}{\rho U^2} = \frac{P}{P_{\infty}} \ \ .
\end{equation}
Note that $\tilde{t}$ and $\mathscr{R}$ are not completely dimensionless due to fractality. Additionally, we employ the formula in Eq. \eqref{fractal_volume_surface} to account for the bulks and boundaries in fractal dimensions. The total time required for a solid sphere at melting temperature to complete {\color{black} liquify} generalizes Eq. \eqref{melt_time_3D} to the following expression:
\begin{equation}
\tilde{\tau}_{\text{m}} (\mathscr{D}_\text{v},\mathscr{D}_\text{s}) =  \frac{\displaystyle \pi^{\mathscr{D}_{\text{v}}-\mathscr{D}_{\text{s}}-1/2} \mathscr{D}_{\text{v}} \Gamma \left( \frac{\mathscr{D}_{\text{s}}+1}{2}\right)}{\displaystyle 4 \left( \mathscr{D}_{\text{v}} - \mathscr{D}_{\text{s}} \right)
 \left( -\mathscr{D}_{\text{v}} + 2\mathscr{D}_{\text{s}} \right) \Gamma\left(\frac{\mathscr{D}_{\text{v}} - \mathscr{D}_{\text{s}}}{2}\right) \Gamma \left( \frac{\mathscr{D}_{\text{v}}+2}{2}\right)} \ \ .
 \label{melting_time_fractal_dimless}
\end{equation}
{\color{black} For verification}, $\tilde{\tau}_{\text{m}}(3,2)=1/2$, agrees with Eq. \eqref{melt_time_3D}. We give the derivation of this, without nondimensionalization, in Appendix \ref{supp_melt_fractal}.

\begin{figure*}[!htbp]
\includegraphics[width=0.5\textwidth]{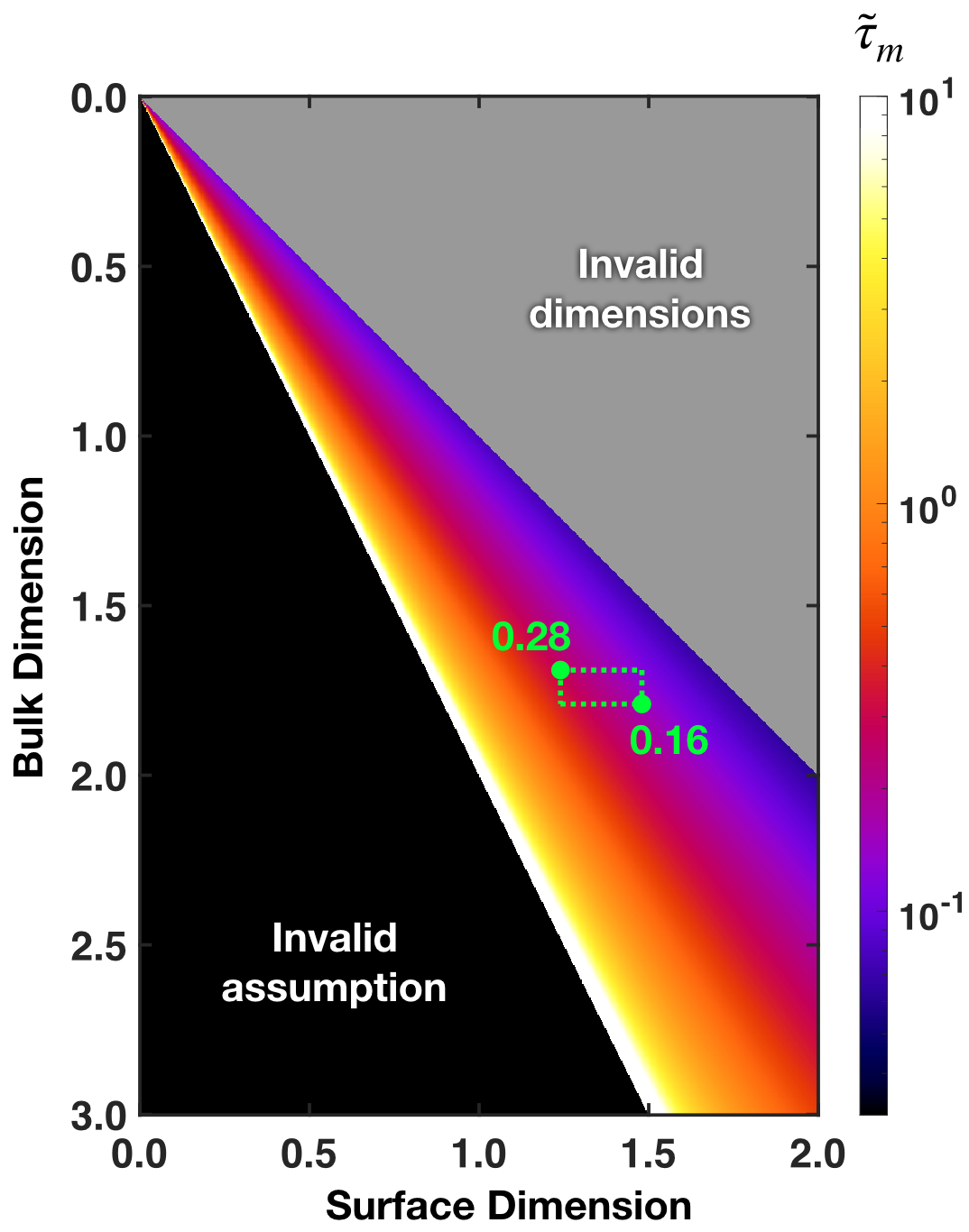}
\caption{\textbf{Thermal melting time in fractal spaces}. The plot for the dimensionless vanishing time $\tilde{\tau}_{\text{m}} (\mathscr{D}_\text{v},\mathscr{D}_\text{s})$ of phase-transitional thermal melting as a function of the volume fractal dimension $\mathscr{D}_\text{v}$ and the surface fractal dimension $\mathscr{D}_\text{s}$, as in Eq. \eqref{melting_time_fractal_dimless}. The green lines locate the fractal dimensionalities of natural soils \cite{gimenez1997fractal}, which give $\tilde{\tau}_{\text{m}} (1.79,1.48) \approx 0.28$ and $\tilde{\tau}_{\text{m}} (1.69,1.24) \approx 0.16$.}
\label{fig01}
\end{figure*}

In Fig. \ref{fig01}, we conduct a numerical investigation of the function $\tilde{\tau}_{\text{m}} (\mathscr{D}_\text{v},\mathscr{D}_\text{s})$  within its valid range of applicability. With dimensionalities as continuous parameters, we discover that the quasi-stationary assumption is applicable for all {\color{black} fractal spaces having the surface dimension $\mathscr{D}_\text{s}$ no less than half of the volume dimension $\mathscr{D}_\text{v}$, i.e. $\mathscr{D}_\text{s} \geq \mathscr{D}_\text{v}/2$}. {\color{black} This can be seen from} Eq. \eqref{melting_time_fractal_dimless}, {\color{black} which shows} $\tilde{\tau}_\text{m} \propto 1/\left(-\mathscr{D}_\text{v}+2\mathscr{D}_\text{s} \right)$. With the same volume dimension, the higher the surface dimension value, the smaller the melting time, which is quite intuitive since this corresponds to larger exposure boundary given the same bulk content. When the surface dimension is too small $\mathscr{D}_\text{s}<\mathscr{D}_\text{v}/2$, the quasi-stationary assumption fails. For fractality similar to that of natural soils, this assumption is applicable, and we can estimate the nondimensionalized melting time to be in between $0.16$ and $0.28$, substantially different than that (of value $0.50$) in three-dimensional Euclidean space.

\section{Hydrodynamic Collapse in Fractal Spaces \label{hyd_frac_main}}

Hydrodynamics in fractal spaces $\left( \mathscr{D}_\text{v},\mathscr{D}_\text{s} \right)$ is very different from that in familiar Euclidean spaces. This deviation is primarily attributed to the incorporation of fractional vector calculus, resulting in the modification of fundamental mathematical operators such as gradient and divergence. Consequently, Eq. \eqref{collapse_ode_main} undergoes a significant transformation, evolving into a notably more intricate expression:
\begin{equation}
\begin{split}
   \left( \mathscr{D}_\text{s}-1 \right) \tilde{R}(\tilde{t}) \frac{d^2}{d\tilde{t}^2}  \tilde{R}(\tilde{t}) & + \left[ \mathscr{D}_\text{s}  \left( \mathscr{D}_\text{s} -1 \right) - \frac12 \right] \left[ \frac{d}{d\tilde{t}} \tilde{R}(\tilde{t}) \right]^2 
   \\
   & + \frac{\displaystyle 2 \mathscr{D}_\text{s}  \Gamma\left(\frac{\mathscr{D}_\text{v} - \mathscr{D}_\text{s}}{2}\right)}{\displaystyle \pi^{(\mathscr{D}_\text{v} - \mathscr{D}_\text{s})/2} \mathscr{R}} \frac1{\tilde{R}^{\mathscr{D}_\text{v} - \mathscr{D}_\text{s}}(\tilde{t})} \frac{d}{d\tilde{t}} \tilde{R}(\tilde{t}) + 1 = 0 \ \ . 
   \end{split}
   \label{fractal_void_eom_main}
\end{equation}
We give the derivation for this, without nondimensionalization, in Appendix \ref{hyd_fractal_supp}. We can check that, for $\left( \mathscr{D}_\text{v},\mathscr{D}_\text{s} \right)=(3,2)$ we get Eq. \eqref{hyd_collapse_3D}, and for $\left( \mathscr{D}_\text{v},\mathscr{D}_\text{s} \right)=(2,1)$ we get Eq. \eqref{plesset_rayleigh_2D}. What {\color{black}is} surprising about this generalization in fractal spaces is that, {\color{black} when} the surface dimension {\color{black} is} small enough, i.e. $\mathscr{D}_\text{s}< 1$, {\color{black} the} shrinkage of void vacuum bubble is not even possible with {\color{black} the fluid initially at rest} $d\tilde{R}(\tilde{t})/d\tilde{t} \Big|_{\tilde{t}=0} = 0$! In such fractal space, as follows from Eq. \eqref{fractal_void_eom_main}, we get the acceleration of the void size {\color{black} at the beginning} to be positive $d^2\tilde{R}(\tilde{t})/d\tilde{t}^2 \Big|_{\tilde{t}=0} > 0$, hence, the void will grow instead of collapse. Since here we are interested in shrinkage, the immediate consequence of this finding is that our simple initial condition is invalid; thus, within the scope of this work, we will not consider $\mathscr{D}_\text{s}< 1$. Beyond the scope of this work, we would like to question what this curious intrinsic instability of fractal fluid might mean to the hydrodynamic fractional continuous models \cite{tarasov2005continuous,tarasov2005possible} for small surface fractal dimensions.

\begin{figure*}[!htbp]
\includegraphics[width=\textwidth]{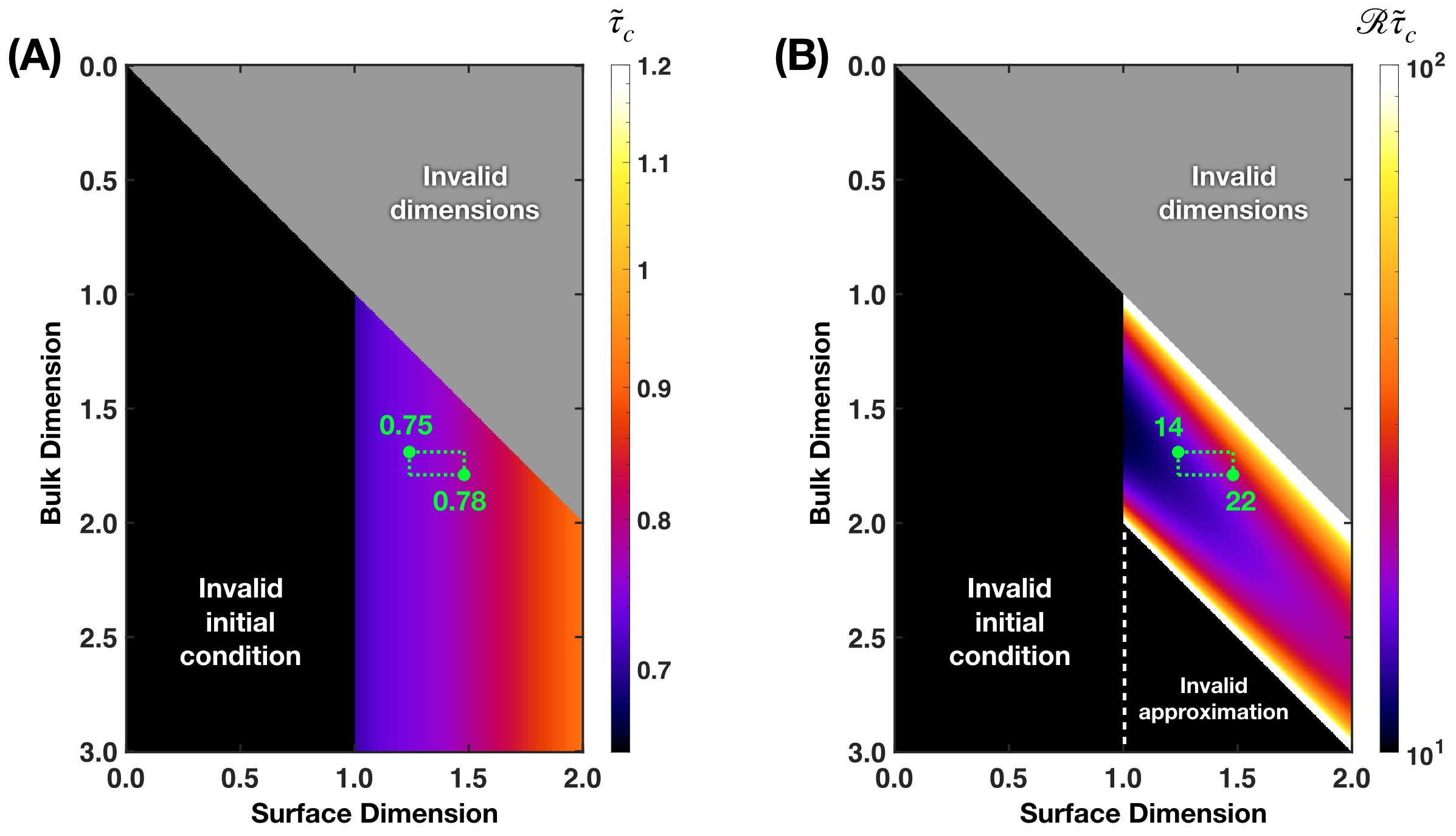}
\caption{\textbf{Hydrodynamic collapse time in fractal spaces, in inviscid and Stokes regimes}. \textbf{(A)} Inviscid regime $\mathscr{R} \rightarrow \infty$. The plot for the dimensionless vanishing time $\tilde{\tau}_{\text{c}} (\mathscr{D}_\text{v},\mathscr{D}_\text{s})$ of hydrodynamic void collapse as a function of the volume fractal dimension $\mathscr{D}_\text{v}$ and the surface fractal dimension $\mathscr{D}_\text{s}$, as in Eq. \eqref{fractal_inviscid_main_eq}. The green lines locate the fractal dimensionalities of natural soils \cite{gimenez1997fractal}, which give $\tilde{\tau}_{\text{c}} (1.79,1.48) \approx 0.78$ and $\tilde{\tau}_{\text{c}} (1.69,1.24) \approx 0.75$. \textbf{(B)} Stokes regime $\mathscr{R} \rightarrow 0$. The plot for $\mathscr{R} \tilde{\tau}_{\text{c}} (\mathscr{D}_\text{v},\mathscr{D}_\text{s})$ as in Eq. \eqref{fractal_stokes_main_eq}. The green lines locate the fractal dimensionalities of natural soils \cite{gimenez1997fractal}, which give $\mathscr{R} \tilde{\tau}_{\text{c}} (1.79,1.48) \approx 22$ and $\mathscr{R} \tilde{\tau}_{\text{c}} (1.69,1.24) \approx 14$.}
\label{fig02}
\end{figure*}

\subsection{Inviscid Flow \label{inviscid_sec}}

Inviscid flow corresponds to inertial-dominated hydrodynamics, which means the effective Reynolds number is extremely large $\mathscr{R} \rightarrow \infty$. In this limit, Eq. \eqref{fractal_void_eom_main} becomes:
\begin{equation}
   \left( \mathscr{D}_\text{s}-1 \right) \tilde{R}(\tilde{t}) \frac{d^2}{d\tilde{t}^2}  \tilde{R}(\tilde{t}) + \left[ \mathscr{D}_\text{s}  \left( \mathscr{D}_\text{s} -1 \right) - \frac12 \right] \left[ \frac{d}{d\tilde{t}} \tilde{R}(\tilde{t}) \right]^2 
   + 1 \approx 0 \ \ , 
\end{equation}
and the time it takes for the vacuum void to get collapsed completely is given by:
\begin{equation}
\begin{dcases}
     \mathscr{D}_\text{s} \geq \frac{1+\sqrt{3}}{2} \ &: \ \tilde{\tau}_{\text{c}} (\mathscr{D}_\text{v},\mathscr{D}_\text{s}) \Big|_{\mathscr{R} \rightarrow \infty} = \frac{\displaystyle \left(\frac{\varphi \pi}{2} \right)^{1/2} \Gamma \left( \frac12 + \frac1\psi \right)}{\displaystyle \psi \Gamma \left( 1 + \frac1\psi \right)}
      \\[10pt]
      1 \leq \mathscr{D}_\text{s} < \frac{1+\sqrt{3}}{2} \ &: \ \tilde{\tau}_{\text{c}} (\mathscr{D}_\text{v},\mathscr{D}_\text{s}) \Big|_{\mathscr{R} \rightarrow \infty} = \frac{\displaystyle \left(\frac{-\varphi \pi}{2} \right)^{1/2} \Gamma \left( \frac1{-\psi} \right)}{\displaystyle -\psi \Gamma \left( \frac12 +\frac1{-\psi} \right)}
      \end{dcases} \ \ ,
\label{fractal_inviscid_main_eq}
\end{equation}
where the coefficients only depend on the surface fractal dimension:
\begin{equation}
\varphi (\mathscr{D}_\text{s}) = 2 \mathscr{D}_\text{s}\left( \mathscr{D}_\text{s}-1 \right) - 1 \ \ , \ \ \psi(\mathscr{D}_\text{s}) = \frac{\varphi(\mathscr{D}_\text{s})}{\mathscr{D}_\text{s}-1} \ \ .
\label{fractal_hyd_collapse_eq}
\end{equation}
{\color{black} We find that the collapsing time $\tilde{\tau}_{\text{c}}$ is a function of the surface dimension $\mathscr{D}_\text{s}$ and independent of the volume dimension $\mathscr{D}_\text{v}$. Note that the volume dimension $\mathscr{D}_\text{v}$ is constrained within the surface dimension $\mathscr{D}_\text{s}$ and the physical space dimensionality $3$.} {\color{black} For verification}, $\tilde{\tau}_{\text{c}}(3,2)=0.91$, agrees with Eq. \eqref{collapsing_time_3D_main}. Our finding also agree with hypersphere collapse \cite{klotz2013bubble}, which corresponds to the special cases $\left( \mathscr{D}_\text{v},\mathscr{D}_\text{s}\right) = (N,N-1)$ with interger-dimensionalities $N \in \mathbb{N}$. In Appendix \ref{supp_fractal_inviscid}, we show how these results can be derived. 

In Fig. \ref{fig02}A, we show our numerical findings about the function $\tilde{\tau}_{\text{c}} (\mathscr{D}_\text{v},\mathscr{D}_\text{s})$  within its valid range of applicability. For inviscid flow, only surface dimension matters, and the higher it is the longer the shrinkage will take. For fractality similar to that of natural soils \cite{gimenez1997fractal}, our initial condition is still valid, and we can estimate the nondimensionalized vanish time of the vacuum void to be in between $0.75$ and $0.78$. It is curious that, right inside this range of dimensionality $\mathscr{D}_\text{s} \in [1.24,1.48]$, $\tilde{\tau}_{\text{c}}$ has a non-trivial and subtle feature which can be seen clearly by investigating the derivative $d\tilde{\tau}_{\text{c}}/d\mathscr{D}_\text{s}$ -- see Fig. \ref{figS01} in Appendix \ref{supp_fractal_inviscid}.  

\subsection{Stokes Flow \label{stokes_flow_sec}}

Stokes flow corresponds to viscosity-dominated hydrodynamics, which means the effective Reynolds number is extremely small $\mathscr{R} \rightarrow 0$. In this limit, we use the following approximation for Eq. \eqref{fractal_void_eom_main}:
\begin{equation}
   \frac{\displaystyle 2 \mathscr{D}_\text{s}  \Gamma\left(\frac{\mathscr{D}_\text{v} - \mathscr{D}_\text{s}}{2}\right)}{\displaystyle \pi^{(\mathscr{D}_\text{v} - \mathscr{D}_\text{s})/2} \mathscr{R}} \frac1{\tilde{R}^{\mathscr{D}_\text{v} - \mathscr{D}_\text{s}}(\tilde{t})} \frac{d}{d\tilde{t}} \tilde{R}(\tilde{t}) + 1 \approx 0 \ \ ,
\label{superviscos_approx}
\end{equation}
and the vanishing time of the bubble can be estimated with:
\begin{equation}
\mathscr{R} \tilde{\tau}_{\text{c}}(\mathscr{D}_\text{v},\mathscr{D}_\text{s}) \Big|_{\mathscr{R} \rightarrow 0} = \frac{\displaystyle 2 \mathscr{D}_\text{s} \Gamma\left(\frac{\mathscr{D}_\text{v} - \mathscr{D}_\text{s}}{2}\right)}{\displaystyle \pi^{(\mathscr{D}_\text{v} - \mathscr{D}_\text{s})/2} \left( 1 + \mathscr{D}_\text{s} -\mathscr{D}_\text{v} \right)} \ \ .
\label{fractal_stokes_main_eq}
\end{equation}
In Appendix \ref{supp_fractal_stokes}, we show how this answer can be derived.

In Fig. \ref{fig02}B, we show our numerical findings about the function $\mathscr{R} \tilde{\tau}_{\text{c}} (\mathscr{D}_\text{v},\mathscr{D}_\text{s})$  within its valid range of applicability. Note that we need to use $\mathscr{R} \tilde{\tau}_{\text{c}}$, not $\tilde{\tau}_{\text{c}}$, to represent the behavior in this regime, so that we can factor out the ever-increasing contribution of viscosity to the collapsing time. Perhaps the most curious feature of the collapsing time, observed when viscosity dominates the hydrodynamics and dimensionalities become parameters, is the existence of a minimum for a fixed volume dimension value $\mathscr{D}_\text{v}$ when the surface dimension $\mathscr{D}_\text{s}$ is somewhere in between $\mathscr{D}_\text{v}-1$ and $\mathscr{D}_\text{v}$, which is the case for fractality similar to natural soils \cite{gimenez1997fractal}. To illustrate, let us take the volume dimension as $\mathscr{D}_{\text{v}}=2$. For surface dimensions of $\mathscr{D}_{\text{s}}=1.1$ and $1.9$, the calculated collapsing times are approximately $\mathscr{R} \tilde{\tau}_{\text{c}} \approx 26$ and $78$, respectively. In the middle of this range, for $\mathscr{D}_{\text{s}}=1.5$, we can estimate a notably faster void collapse with $\mathscr{R} \tilde{\tau}_{\text{c}} \approx 16$. The origin of this minimization can be traced back to the coupling between viscosity and fractality of space in the Laplacian term of the hydrodynamic Navier-Stokes equation. For $\mathscr{D}_{\text{s}} < \mathscr{D}_{\text{v}} -1 $, our approximation Eq. \eqref{superviscos_approx} quantitatively becomes invalid, but qualitatively it still captures the behavior that the void collapsing time of highly-viscous fluid in such fractal spaces becomes extremely large. We will see more about this, as how such emerges when the effective Reynolds number $\mathscr{R}$ increases, in Section \ref{general_viscous_flow}.

\subsection{General Viscous Flow \label{general_viscous_flow}}

We can solve Eq. \eqref{fractal_void_eom_main} numerically, to investigate the general influence of viscosity to hydrodynamic collapse process. We use the {\it ode45} package of MatLab 2023a \cite{MATLAB} to carry out this task, so that we can enforce variables to always stay non-negative. Analytical features of the surface $\tilde{\tau}_\text{c} \left( \mathscr{D}_\text{v},\mathscr{D}_\text{s}\right)$ in the extreme limits of low-viscosity $\mu \rightarrow 0$ (which means $\mathscr{R}\rightarrow \infty$) and high-viscosity $\mu \rightarrow \infty$ (corresponded to $\mathscr{R}\rightarrow 0$) has been analyzed in Section \ref{inviscid_sec} and Section \ref{stokes_flow_sec}, thus we expect our numerical results to not only confirm our previous findings but also map out the transitional progression (as we decreases $\mathscr{R}$) between these two regimes.

\begin{figure*}[!htbp]
\includegraphics[width=\textwidth]{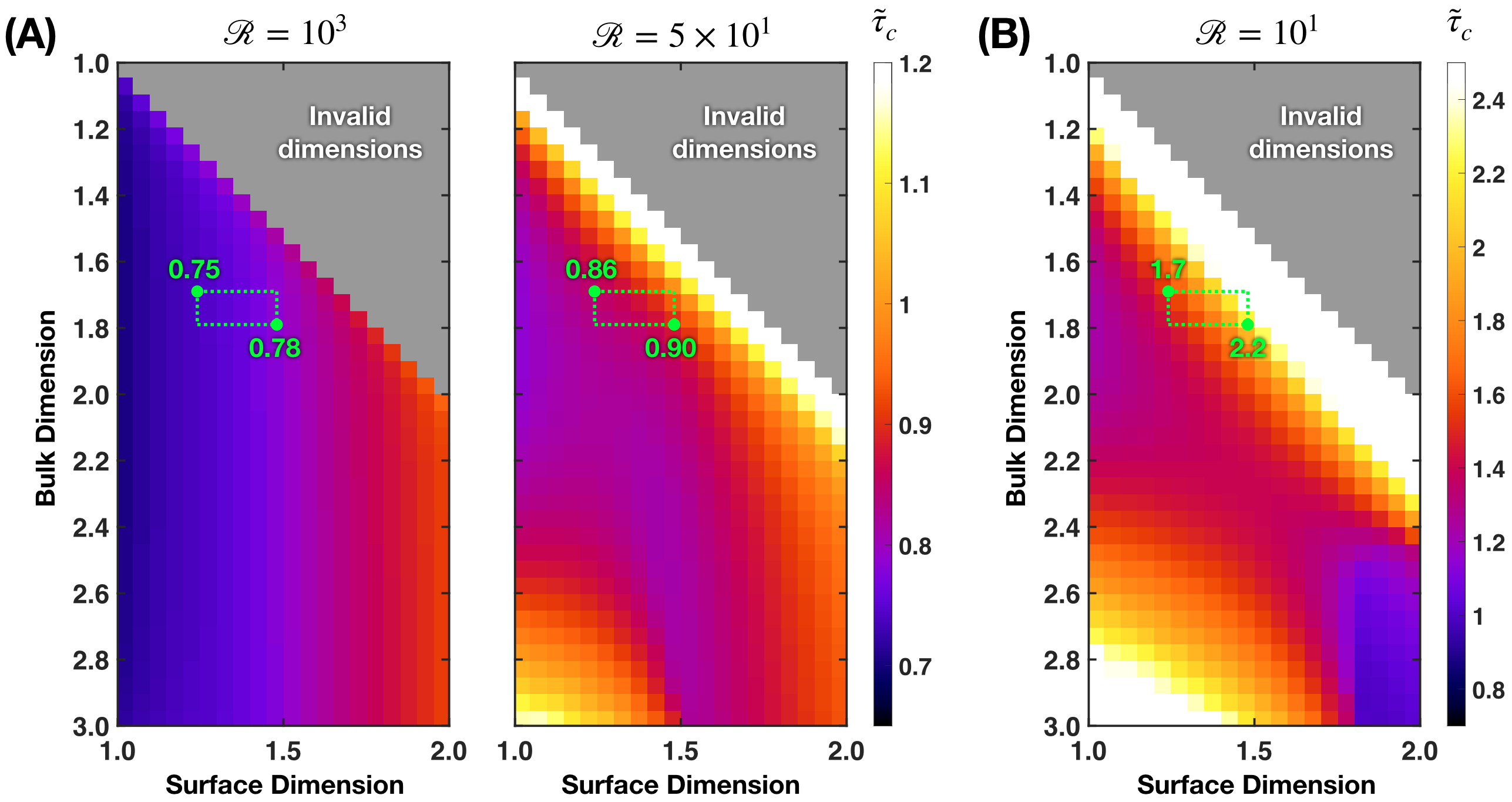}
\caption{\textbf{The influence of viscosity on the collapsing time of vacuum void.} The green lines specify the dimensionalities of natural soils \cite{gimenez1997fractal} and the green values indicate the corresponding $\tilde{\tau}_{\text{c}}$. \textbf{(A)} Increasing viscosity $\mu$ decreases the effective Reynolds number $\mathscr{R}$. For a large value $\mathscr{R}=10^3$, the nondimensionalized collapsing time $\tilde{\tau}_{\text{c}}$ is about the same as shown in Fig. \ref{fig02}A. For smaller value $\mathscr{R}=5 \times 10^1$, $\tilde{\tau}_{\text{c}}$ becomes very large near the corner $\left( \mathscr{D}_\text{v},\mathscr{D}_\text{s}\right) = (3,1)$ and near the edge $\mathscr{D}_\text{v}=\mathscr{D}_\text{s}$. \textbf{(B)} $\tilde{\tau}_{\text{c}} \left( \mathscr{D}_\text{v},\mathscr{D}_\text{s}\right)$ for $\mathscr{R}=10^1$ looks similar to Fig. \ref{fig02}B, in which the invalid approximation corner there corresponds to the corner of slow collapse here.}
\label{fig03}
\end{figure*}

Fig. \ref{fig03}A showcases the difference of the collapsing time surface $\tilde{\tau}_\text{c} \left( \mathscr{D}_\text{v},\mathscr{D}_\text{s}\right)$ between high and low effective Reynolds number. As $\mathscr{R}$ becomes smaller, the effect of the Laplacian term $\propto 1/\mathscr{R}$ that couples viscosity with fractality in Eq. \eqref{fractal_void_eom_main} dominates. The influence of such contributions can be seen from Eq. \eqref{fractal_stokes_main_eq}: the collapsing time diverges at the edges $\mathscr{D}_\text{v}=\mathscr{D}_\text{s}$ and $\mathscr{D}_\text{v}-1=\mathscr{D}_\text{s}$. We can see how these emerges, as in $\tilde{\tau}_\text{c} \left( \mathscr{D}_\text{v},\mathscr{D}_\text{s}\right)$ the edge $\mathscr{D}_\text{v}=\mathscr{D}_\text{s}$ becomes sharper and the corner near $\left( \mathscr{D}_\text{v},\mathscr{D}_\text{s}\right)=(3,1)$, which corresponds to invalid approximation in Fig. \ref{fig02}B, gradually rises up. Consider $\mathscr{D}_\text{v}=3$ on the right-plot $\mathscr{R}=5\times 10^1$ in Fig. \ref{fig02}A, a minimum exists at around $\mathscr{D}_\text{s} \approx 1.5$, far away from the end points $\mathscr{D}_\text{s}=1$ and $\mathscr{D}_\text{s}=2$.

More details on the transitional progression in Fig. \ref{fig03}A can be seen in Appendix \ref{slowmo_app}. We show $\tilde{\tau}_\text{c} \left( \mathscr{D}_\text{v},\mathscr{D}_\text{s}\right)$ when the effective Reynolds number is very small $\mathscr{R}=10^1$ in Fig. \ref{fig03}B. The features are now very similar to those in Fig. \ref{fig03}B, validating our approximation Eq. \eqref{fractal_stokes_main_eq}.

\section{Discussion}

 {\color{black} We have investigated} the vanishing dynamics, an important category of physical transport phenomenon observed in both heat and mass transfer processes i.e. with thermal melting \cite{mccue2008classical} and hydrodynamic collapse  \cite{walton1984sonoluminescence,margulis1995sonochemistry,young1989cavitation,barber1991observation}, describing gradual disappearances within defined environments over time. Though these manifestations have been extensively studied in familiar Euclidean spaces, examining them in fractal spaces -- with dimensions as continuous parameters -- brings forth not only generalizations but also a wealth of surprises and insights. For a phase-transitional thermal melting, we extend the quasi-stationary approximation for estimating liquefaction time to spaces in which the volume dimension exceeds the surface dimension by at least two-fold. For a hydrodynamic collapse of vacuum void, we reveal an emergent mass transport optimization through the interplay of fluid viscosity and surface fractal dimension, emphasizing engineering potential of fractality in hydrodynamics. The range of volume and surface fractal dimensions considered in our work encompasses those observed in natural soils \cite{gimenez1997fractal}.

 Fractals, while appearing complex, are created simply by repeating the same patterns. Mother nature has employed fractal designs for eons, and it is only in recent times that humans have started emulating these natural structures for innovative device design and development. The investigation of transport phenomena such as heat transfer and fluid flow within fractal spaces not only deepens our scientific understanding but also translates into tangible advancements in various industries. By capitalizing on the unique attributes of fractal geometries, engineers and scientists are poised to revolutionize processes and technologies across sectors ranging from energy to environmental management and nanomedicine \cite{brambila2017fractal,fraser2014distillation,ochoa2015optimization,dupain2006optimal,sundberg2018optimal,shang2022thermal,sirine2022numerical,jackson2014environmental,yu2008analysis,balankin2012map,nandlal2019drained,chen2023biomimetic,wang2019interactions,pippa2013ubiquitous,samioti2019effect,alexiou2023fractal}. 

{\color{black} Beyond pure physics, similar analyses of PDEs using Eq. \eqref{fractal_grad} and Eq. \eqref{fractal_div} can extend into the realm of biology. For example,} understanding how {\color{black} active fluids (e.g. microorganisms and cells)} grow, propagate \cite{mattingly2019rule,narla2021traveling,phan2023direct}, {\color{black} exchange information, and perform works \cite{ro2022model,nguyen2023remark}} within fractal environments offers a more accurate depiction of their collective behavior in real-world natural habitats \cite{phan2020bacterial}. This stands in contrast to the setting commonly used wet labs, where organisms are often studied on featureless agar plates. The distinct geometrical features of fractal spaces introduce a range of complexities that organisms must navigate. These intricacies provide both challenges and potential advantages, ultimately influencing their ability to adapt and flourish in such environments. This research avenue enables a deeper understanding of the nuanced interplay between organisms and their complex surroundings.  Overall, there exists a large landscape of possibilities for research in the realm of transport phenomena in fractal spaces.

\appendix 

\section{The Laplacian \label{Lap}}

The Laplacian $\Delta = \vec{\nabla}.\vec{\nabla}$ acting on the radial scalar function $S(r)$ and on the radial vector function $\vec{V}(r) = V(r) \hat{r}$, calculated directly by applying Eq. \eqref{fractal_grad} and Eq. \eqref{fractal_div}, gives us the followings:
\begin{equation}
\begin{split}
\Delta S(r) &= A\left( \mathscr{D}_{\text{v}}, \alpha_r \right) \left[ \frac1{r^{2\alpha_r - 2}} \partial_r^2 + \frac{\mathscr{D}_{\text{v}} - 2\alpha_r + 1}{r^{2\alpha_r - 1}} \partial_r \right] S(r) \ \ ,
\\
\Delta \vec{V}(r) & = A\left( \mathscr{D}_{\text{v}}, \alpha_r \right) \left[ \frac1{r^{2\alpha_r - 2}} \partial_r^2 +\frac{\mathscr{D}_{\text{v}} - 2\alpha_r + 1}{r^{2\alpha_r - 1}} \partial_r - \frac{(\mathscr{D}_{\text{v}}-\alpha_r) \alpha_r}{r^{2 \alpha_r}} \right] V(r) 
 \ \ ,
\end{split}
\label{fractal_laplacian}
\end{equation}
where we need the prefactor:
\begin{equation}
A\left( \mathscr{D}_{\text{v}}, \alpha_r \right) = \frac{\displaystyle \Gamma\left(\frac{\alpha_r}{2}\right) \Gamma\left( \frac{\mathscr{D}_{\text{v}}}{2}\right)}{\displaystyle \pi^{\alpha_r - 1/2} \Gamma\left( \frac{\mathscr{D}_{\text{v}} -\alpha_r +1}{2}\right)} \ \ .
\label{the_A}
\end{equation}
These are in agreement with what has been reported in \cite{tarasov2014flow}.

\section{Vanishing Dynamics in Three-Dimensional Euclidean Space Revisit}

\subsection{Thermal Melting \label{therm_melt_3D}}

Consider a solid sphere of melting temperature $T_{\text{m}}$ is submerged deeply inside a liquid medium at a hotter ambient temperature $T_{\infty} > T_{\text{m}}$, in physical three-dimensional Euclidean space $\left( \mathscr{D}_{\text{v}}, \mathscr{D}_{\text{s}} \right) =(3,2)$. For extreme regime i.e. the environment outside the melting sphere having negligible heat capacity $C_{\text{l}} \rightarrow 0$, we can assume the total 
 incoming heat flux
 \begin{equation}
     \vec{J}(r,t) = J(r,t) \hat{r} = \mathcal{A}(r) \vec{j}(r,t) \ \ , \ \ \vec{j}(r,t) =  - \kappa \vec{\nabla}  T(r,t) = - \kappa \partial_r T(r,t) \hat{r}
     \label{heat_flux}
 \end{equation}
 to be radially-uniform as the temperature $T$ are at quasi-stationary for all time $t$:
\begin{equation}
 \partial_t T(r,t) \propto \Delta T(r,t) \rightarrow 0 \ \ \Longrightarrow \ \ \vec{\nabla}  \vec{j}(r,t) \propto \partial_r J(r,t) \rightarrow 0 \ \ ,
 \label{heat_flux_radial_uniform}
\end{equation}
where $\kappa$ is the heat conductivity of the surrounding liquid and $\mathcal{A}(r)=4\pi r^2$ is the surface area of a sphere having radius $r$. Here we utilize rotational symmetry $T(\vec{r},t) \rightarrow T(r,t)$ around the ice sphere centered at the origin $r=0$, and apply this equation in the water region $r \in \left[ R(t), \infty \right)$. From Eq. \eqref{heat_flux} and Eq. \eqref{heat_flux_radial_uniform}, we can relate the flow of heat with the temperature profile by treating $J(r,t)$ as a constant:
\begin{equation}
J(r,t) = \mathcal{A}(r) \left[ - \kappa \partial_r T(r,t)  \right] \ \ \Longrightarrow \ \ J(r,t) = -\kappa \frac{\displaystyle T(r',t)\Big|^{r'=\infty}_{r'=r}}{\displaystyle \int^{\infty}_r dr' \mathcal{A}^{-1}(r')} \ \ .
\label{heat_flux_constant}
\end{equation}
 With the temperature near the surface of the melting ice sphere at radius $R(t)$ to be $T(r)\Big|_{r=R(t)} = T_{\text{m}}$ and the temperature at far-away infinity $r\rightarrow \infty$ to be $T_{\infty}$, we can arrive at the total heat that melts the solid sphere per unit time:
\begin{equation}
\begin{split}
\frac{d}{dt} Q & \equiv - L \frac{d}{dt} \mathcal{V}(r)\Big|_{r=R(t)} 
\\
& = -J(r,t)\Big|_{r=R(t)} = \kappa \left[ 4\pi 
 R(t) \right] \left( T_{\infty} - T_{\text{m}} \right) \ \ ,
\end{split}
\end{equation}
where $\mathcal{V}(r)=4\pi r^3/3$ is the volume of a sphere having radius $r$. We then obtain the equation describing the size $R(t)$ of the solid sphere:
\begin{equation}
- L \frac{d}{dt} \left[ \frac{4\pi}{3} R^3(t) \right]  =  \kappa \left[ 4\pi 
 R(t) \right] \left( T_{\infty} - T_{\text{m}} \right) \ \ \Longrightarrow \ \ \frac{d}{dt} R(t) = -\frac{\kappa \left( T_{\infty}-T_{\text{m}}\right)}{L R (t)} \ \ ,
\end{equation}
which we can solve analytically for the total melting time:
\begin{equation}
\tau_{\text{m}} = \int_0^{R(0)} dR' \frac{LR'}{\kappa \left( T_{\infty}-T_{\text{m}}\right)} = \frac12 \left[ \frac{LR^2(0)}{ \kappa \left( T_{\infty}-T_{\text{m}}\right)} \right] \ \ .
\label{melting_time_3D}
\end{equation}
This reveals a quadratic-dependent on the initial value of the solid sphere radius $\tau_{\text{m}} \propto R^2(0)$.

\subsection{Hydrodynamic Collapse \label{hyd_collapse_3D}}

Consider the hydrodynamic collapse of a spherical vacuum bubble in physical three-dimensional Euclidean space $\left( \mathscr{D}_{\text{v}}, \mathscr{D}_{\text{s}} \right) =(3,2)$, given the ambient pressure $P_\infty$ and assume negligible surface tension. Let us start with the incompresibility condition for spherical collapse, which allows us to write the radial velocity field every where $\vec{v}(r,t) = v(r,t) \hat{r}$ via how the cavity (centered at the origin) of size $R(t)$ changes with time:
\begin{equation}
\begin{split}
\vec{\nabla} .\vec{v}(r,t) = 0 \ \ &\Longrightarrow \ \ v(r,t) \propto r^{-2} \ \ ,
\\
v(r,t) \Big|_{r=R(t)} = \frac{d}{dt} R(t) \ \ &\Longrightarrow \ \ v(r,t) = \left[ \frac{R(t)}{r} \right]^2 \frac{d}{dt} R(t) \ \ .
\end{split}
\label{vec_field}
\end{equation}
The Navier-Stokes equation describing an incompressible fluid of mass density $\rho$ is given by:
\begin{equation}
\partial_t \vec{v}(r,t) + \left[ \vec{v} (r,t) . \vec{\nabla} \right] \vec{v}(r,t) = -\frac1\rho \vec{\nabla} P(r,t) + \frac1\rho \vec{\nabla}.\sigma(r,t) \ \ ,
\label{NS_eq}
\end{equation}
where $P(r,t)$ is the pressure and $\sigma(r,t)$ is the deviatoric stress-tensor \cite{stokes2007theories}:
\begin{equation}
\frac1\rho \sigma = \mu \left[ \vec{\nabla}\vec{v} + (\vec{\nabla}\vec{v} )^{T} \right] \ \ \Longrightarrow \ \ \frac1\rho \vec{\nabla}.\sigma = \mu \Delta \vec{v}(r,t) \ \ .
\label{deviatoric}
\end{equation}
Here $\mu$ is the kinematic shear-viscosity. We then arrive at:
\begin{equation}
\partial_t v (r,t) + v(r,t) \partial_r v(r,t) = - \frac1\rho \partial_r P(r,t) + \mu \left[ \partial_r^2 + \frac{2}{r} \partial_r - \frac{2}{r^2} \right] v(r,t) \ \ .
\end{equation}
Let us radially integrate this equation from $r=R(t)$ to $ r\rightarrow \infty$: 
\begin{equation}
\begin{split}
&-\int_{R(t)}^{\infty} dr \left[ \partial_t v(r,t) \right] + \mu \int_{R(t)}^{\infty} dr \left[ \partial_r^2 + \frac{2}{r} \partial_r - \frac{2}{r^2} \right] v(r,t)
\\
& \ \ \ \ \ \ \ \ \ \ \ \ \ = \left[ \frac12 v^2(r,t) + \frac1\rho P(r,t) \right] \Bigg|^{r\rightarrow \infty}_{r=R(t)}  \ \ .
\end{split}
\label{r_integrate}
\end{equation}
We arrange the terms this way, so that on the second-line we can see the Bernoulli's principle when the first-line goes to 0 i.e. stationary flow profile $\partial_t v \rightarrow 0$ at inviscid limit $\mu \rightarrow 0$. At far-away infinity $r\rightarrow \infty$, the pressure is $P=P_\infty$. At the surface of the spherical void $r=R(t)$, the pressure is equal to the viscous normal-stress from Eq. \eqref{deviatoric}:
\begin{equation}
\frac1\rho P(r)\Big|_{r=R(t)} = \frac1\rho \sigma_{rr}(r,t) \Big|_{r=R(t)} = 2 \mu \partial_r \left\{ \left[ \frac{R(t)}{r}\right]^2 \frac{d}{dt} R(t) \right\} \Bigg|_{r=R(t)} = - \frac{4\mu}{R(t)} \frac{d}{dt} R(t) \ \ .
\end{equation}
With the velocity field as in Eq. \eqref{vec_field}, we obtain:
\begin{equation}
\begin{split}
    -\frac1{R(t)} \frac{d}{dt} \left[ R^2(t) \frac{d}{dt} R(t) \right] = - \frac12 \left[\frac{d}{dt} R(t) \right]^2 - \frac1\rho \left[ -\frac{4\mu}{R(t)} \frac{d}{dt} R(t) \right] + \frac1\rho P_{\infty}
    \\
    \Longrightarrow \ \ R(t) \frac{d^2}{dt^2} R(t) + \frac32 \left[\frac{d}{dt} R(t) \right]^2 + \frac{4\mu}{R(t)}  \frac{d}{dt} R(t) + \frac1\rho P_{\infty}= 0 \ \ ,
\end{split}
\end{equation}
note that the viscous term $\mu \int dr ...$ vanishes when we do the evaluation for the first-line. In the inviscid limit $\mu \rightarrow 0$, we can rewrite the second-differential term $d^2/dt^2$ (which corresponds to bubble size acceleration):
\begin{equation}
\frac{1}{2} R \frac{\displaystyle d \left( \frac{d}{dt} R \right)^2}{dR} + \frac32 \left(\frac{d}{dt} R \right)^2 + \frac1\rho P_{\infty} \approx 0 \ \ \Longrightarrow \ \ \frac{\displaystyle d \left( \frac{d}{dt} R \right)^2}{\displaystyle \left(\frac{d}{dt} R \right)^2 + \frac{2 P_{\infty}}{3\rho} } = - 3 \frac{dR}{R} \ \ .
\end{equation}
Consider that initially the void has size $R(0)$ and we assume it can have no velocity $dR(t)/dt \Big|_{t=0} = 0$, then we can integrate both sides of the above equation from time $0$ to $t$:
\begin{equation}
\begin{split}
    & \ln \left\{ \frac{\displaystyle \left[\frac{d}{dt} R(t) \right]^2 + \frac{2 P_{\infty}}{3\rho}}{\displaystyle  \frac{2 P_{\infty}}{3\rho}} \right\} = - 3 \ln \left[ \frac{R(0)}{R(t)} \right] 
    \\
    & \ \ \ \ \ \ \ \ \ \ \ \  
 \Longrightarrow \ \ \frac{d}{dt} R(t) = -\left( \frac{2P_{\infty}}{3\rho} \right)^{1/2} \left\{ \left[ \frac{R(0)}{R(t)} \right]^3 - 1 \right\}^{1/2} \ \ .
    \end{split}
\end{equation}
Hence the void collapsing time can be found to be:
\begin{equation}
\tau_{\text{c}} = \left( \frac{3\rho}{2P_{\infty}} \right)^{1/2}\int_0^{R(0)} dR' \left\{ \left[ \frac{R(0)}{R'} \right]^3 - 1 \right\}^{-1/2} 
= \frac{\displaystyle \left(\frac{3\pi}{2}\right)^{1/2} \Gamma \left( \frac56 \right)}{\displaystyle 3 \Gamma \left(\frac43 \right)}
\left( \frac{\rho}{P_{\infty}} \right)^{1/2} R(0) \ \ .
\label{collapsing_time_3D}
\end{equation}
This reveals a linear-dependent on the initial size of the sphere vacuum void $\tau_{\text{c}} \propto R(0)$. 

\section{Analytical Investigation of Thermal Melting in Fractal Space \label{supp_melt_fractal}}

Similar to Eq. \eqref{heat_flux}, as follows from how the gradient operator is defined Eq. \eqref{fractal_grad} in a general fractal space $\left(\mathscr{D}_\text{v},\mathscr{D}_\text{s}\right)$, the total heat flux can be calculated as:
\begin{equation}
J(r,t) = - \kappa \frac{\displaystyle 2\pi^{(\mathscr{D}_{\text{v}}-2\alpha_r+1)/2} \Gamma\left(\frac{\alpha_r}{2}\right) }{\displaystyle \Gamma \left( \frac{\mathscr{D}_{\text{v}}-\alpha_r+1}{2}\right)} r^{\mathscr{D}_{\text{v}}-2\alpha_r+1}  \partial_r T(r,t)  \ \ ,
\end{equation}
where $\alpha_r = \mathscr{D}_\text{v} - \mathscr{D}_\text{s}$ is the fractal dimensionality of the radial direction and here we have used the surface area of a sphere radius $r$ in this space to be:
\begin{equation}
\mathcal{A}(r) = \frac{\displaystyle 2\pi^{(\mathscr{D}_{\text{s}}+1)/2}}{\displaystyle \Gamma \left( \frac{\mathscr{D}_{\text{s}}+1}{2}\right)} r^{\mathscr{D}_{\text{s}}} =  \frac{\displaystyle 2\pi^{(\mathscr{D}_{\text{v}}-\alpha_r+1)/2}}{\displaystyle \Gamma \left( \frac{\mathscr{D}_{\text{v}}-\alpha_r+1}{2}\right)} r^{\mathscr{D}_{\text{v}}-\alpha_r} \ \ .
\end{equation}
We can then relate the flow of heat with the temperature profile in the same way as Eq. \eqref{heat_flux_constant}:
\begin{equation}
-J(r,t)\Big|_{r=R(t)} = \frac{\displaystyle 2\pi^{(\mathscr{D}_{\text{v}}-2\alpha_r+1)/2} 
 \left( \mathscr{D}_{\text{v}} - 2\alpha_r \right) \Gamma\left(\frac{\alpha_r}{2}\right) }{\displaystyle \Gamma \left( \frac{\mathscr{D}_{\text{v}}-\alpha_r+1}{2}\right)} \kappa \left( T_{\infty} - T_{\text{m}} \right) R^{ \mathscr{D}_{\text{v}} - 2\alpha_r }(t)  \ \ .
\end{equation}
This total flux of heat comes toward the solid sphere, so we have to equate it with the released heat due to melting:
\begin{equation}
-L \frac{d}{dt} \mathcal{V}(r) \Big|_{r=R(t)} =  -\frac{\displaystyle \pi^{\mathscr{D}_{\text{v}}/2} \mathscr{D}_{\text{v}}}{\displaystyle \Gamma \left( \frac{\mathscr{D}_{\text{v}}+2}{2}\right)} L R^{\mathscr{D}_{\text{v}}-1}(t) \frac{d}{dt} R(t) \ \ ,
\end{equation}
in which we have used the volume of a sphere radius $r$ in the fractal space to be:
\begin{equation}
\mathcal{V}(r) = \frac{\displaystyle \pi^{\mathscr{D}_{\text{v}}/2}}{\displaystyle \Gamma \left( \frac{\mathscr{D}_{\text{v}}+2}{2}\right)} r^{\mathscr{D}_{\text{v}}}  \ \ .
\end{equation}
We therefore obtain the total melting time of the solid sphere:
\begin{equation}
\begin{split}
\tau_{\text{m}} &= \frac{\displaystyle \pi^{\alpha_r-1/2} \mathscr{D}_{\text{v}}  \Gamma \left( \frac{\mathscr{D}_{\text{v}}-\alpha_r+1}{2}\right)}{\displaystyle 2
 \left( \mathscr{D}_{\text{v}} - 2\alpha_r \right) \Gamma\left(\frac{\alpha_r}{2}\right) \Gamma \left( \frac{\mathscr{D}_{\text{v}}+2}{2}\right)} \int^{R(0)}_0 dR' \frac{L {R'}^{2\alpha_r -1}dR'}{\kappa \left( T_{\infty} - T_{\text{m}} \right)} 
 \\
 &= \frac{\displaystyle \pi^{\alpha_r-1/2} \mathscr{D}_{\text{v}} \Gamma \left( \frac{\mathscr{D}_{\text{v}}-\alpha_r+1}{2}\right)}{\displaystyle 4 \alpha_r 
 \left( \mathscr{D}_{\text{v}} - 2\alpha_r \right) \Gamma\left(\frac{\alpha_r}{2}\right) \Gamma \left( \frac{\mathscr{D}_{\text{v}}+2}{2}\right)} \left[ \frac{L R^{2\alpha_r}(0)}{\kappa \left( T_{\infty} - T_{\text{m}} \right)} \right] \ \ .
 \end{split}
 \label{melting_time_fractal}
\end{equation}
Here we show a power-law dependency, on the solid sphere initial radius with the radial fractal dimensionality, of the melting time  $\tau_{\text{m}} \propto R^{2\alpha_r}(0)$.

\section{Analytical Investigation of Hydrodynamic Collapse in Fractal Space \label{hyd_fractal_supp}}

Similar to Eq.\eqref{vec_field}, as follows from how the divergence operator is defined Eq. \eqref{fractal_div} in a general fractal space $\left(\mathscr{D}_\text{v},\mathscr{D}_\text{s}\right)$, the incompressibility condition gives:
\begin{equation}
\vec{\nabla} \vec{v}(r,t) \propto \left( \frac1{\displaystyle 
 r^{\alpha_r - 1}} \partial_r + \frac{\displaystyle \mathscr{D}_{\text{v}} - \alpha_r}{\displaystyle 
 r^{\alpha_r}} \right) v(r,t) = 0 \ \ \Longrightarrow \ \ v(r,t) = \left[ \frac{R(t)}{r} \right]^{\mathscr{D}_\text{v} - \alpha_r} \frac{d}{dt} R(t) \ \ .
 \label{fractal_vec_field}
\end{equation}
In this space, the Navier-Stokes equation Eq. \eqref{NS_eq} becomes \cite{tarasov2014flow}:
\begin{equation}
\begin{split}
& \partial_t v (r,t) + \frac{\displaystyle \Gamma\left(\frac{\alpha_r}{2}\right)}{\displaystyle \pi^{\alpha_r/2}} \frac1{\displaystyle r^{\alpha_r-1}} v(r,t) \partial_r v(r,t) = 
\\
& \ \ - \frac1\rho \partial_r P(r,t) + \mu A\left( \mathscr{D}_{\text{v}}, \alpha_r \right) \left[ \frac1{r^{2\alpha_r - 2}} \partial_r^2 +\frac{\mathscr{D}_{\text{v}} - 2\alpha_r + 1}{r^{2\alpha_r - 1}} \partial_r - \frac{(\mathscr{D}_{\text{v}}-\alpha_r) \alpha_r}{r^{2 \alpha_r}} \right] v(r,t) \ \ ,
\end{split}
\end{equation}
where we have used Eq. \eqref{fractal_laplacian} and Eq. \eqref{the_A}. Here, use the radial vector field in Eq. \eqref{fractal_vec_field} and radially integrate from $r=R(t)$ to $ r\rightarrow \infty$ as in Eq. \eqref{r_integrate} gives:
\begin{equation}
 -  \frac{\mathscr{D}_\text{v} - \alpha_r-1}{R^{\mathscr{D}_\text{v} - \alpha_r-1}}\frac{d}{dt} \left( R^{\mathscr{D}_\text{v} - \alpha_r} \frac{d}{dt} R \right) = -\frac12 \left( \frac{d}{dt} R\right)^2 + \frac1\rho P \Big|_{r=R}^{r\rightarrow \infty} \ \ .
 \label{size_eom}
\end{equation}
Note that the viscous term $\mu \int ...$ vanishes. At far-away infinity $r\rightarrow \infty$, the pressure is $P=P_\infty$. At the surface of the spherical void $r=R(t)$, the pressure is equal to the viscous normal-stress from Eq. \eqref{deviatoric}:
\begin{equation}
\frac1\rho P(r)\Big|_{r=R} = \frac1\rho \sigma_{rr} \Big|_{r=R} =
 - \frac{\displaystyle 2 (\mathscr{D}_\text{v} - \alpha_r) \Gamma\left(\frac{\alpha_r}{2}\right) \mu}{\displaystyle \pi^{\alpha_r/2} R^{\alpha_r} } \frac{d}{dt} R  \ \ ,
\end{equation}
in which we apply the gradient operator as in Eq. \eqref{fractal_grad}. Put all these together in Eq. \eqref{size_eom}, we get the equation of motion for the void vacuum size:
\begin{equation}
\begin{split}
   \left( \mathscr{D}_\text{v} - \alpha_r-1 \right) R \frac{d^2}{dt^2}  R + \left[ \left( \mathscr{D}_\text{v} - \alpha_r \right) \left( \mathscr{D}_\text{v} - \alpha_r-1 \right) - \frac12 \right] \left( \frac{d}{dt} R\right)^2 &
   \\
   + \frac{\displaystyle 2 (\mathscr{D}_\text{v} - \alpha_r) \Gamma\left(\frac{\alpha_r}{2}\right) \mu}{\displaystyle \pi^{\alpha_r/2} R^{\alpha_r} } \frac{d}{dt} R + \frac1\rho P_{\infty}= 0 & \ \ . 
   \end{split}
   \label{fractal_void_eom}
\end{equation}
For $\left( \mathscr{D}_\text{v},\mathscr{D}_\text{s}\right) = \left(2,1 \right)$, we get Eq. \eqref{plesset_rayleigh_2D}.

\subsection{Inviscid Flow \label{supp_fractal_inviscid}}

Consider the inviscid limit $\mu \rightarrow 0$, Eq. \eqref{fractal_void_eom} becomes:
\begin{equation}
\begin{split}
   \frac{ \mathscr{D}_\text{v} - \alpha_r-1}{2} R \frac{\displaystyle d \left( \frac{d}{dt} R \right)^2}{dR}+ \left[ \left( \mathscr{D}_\text{v} - \alpha_r \right) \left( \mathscr{D}_\text{v} - \alpha_r-1 \right) - \frac12 \right] \left( \frac{d}{dt} R\right)^2 + \frac1\rho P_{\infty} \approx 0  &
   \\
   \Longrightarrow \ \ \frac{\displaystyle d \left( \frac{d}{dt} R \right)^2}{\displaystyle \left( \frac{d}{dt} R \right)^2 + \frac{2P_{\infty}}{\varphi(\mathscr{D}_\text{s}) \rho}} = - \psi(\mathscr{D}_\text{s}) \frac{dR}{R} & \ \ ,
   \end{split} 
\end{equation}
in which the coefficients are:
\begin{equation}
\varphi (\mathscr{D}_\text{s}) = 2 \mathscr{D}_\text{s}\left( \mathscr{D}_\text{s}-1 \right) - 1 \ \ , \ \ \psi(\mathscr{D}_\text{s}) = \frac{\varphi(\mathscr{D}_\text{s})}{\mathscr{D}_\text{s}-1} \ \ .
\label{coeffs}
\end{equation}
Consider that initially the void has size $R(0)$ and we assume it can have no velocity $dR(t)/dt\Big|_{t=0} = 0$, then we can integrate both sides of the above equation from time $0$ to $t$:
\begin{equation}
\begin{split}
    & \ln \left\{ \frac{\displaystyle \left[\frac{d}{dt} R(t) \right]^2 + \frac{2 P_{\infty}}{\varphi \rho}}{\displaystyle  \frac{2 P_{\infty}}{\varphi \rho}} \right\} = - \psi \ln \left[ \frac{R(0)}{R(t)} \right] 
    \\
    & \ \ \ \ \ \ \ \ 
 \Longrightarrow \ \ \begin{dcases}
     \mathscr{D}_\text{s} \geq \frac{1+\sqrt{3}}{2} \ &: \ \frac{d}{dt} R(t) = -\left( \frac{2P_{\infty}}{\varphi \rho} \right)^{1/2} \left\{ \left[ \frac{R(0)}{R(t)} \right]^\psi - 1 \right\}^{1/2}
      \\[10pt]
      1 \leq \mathscr{D}_\text{s} < \frac{1+\sqrt{3}}{2} \ &: \ \frac{d}{dt} R(t) = -\left( \frac{2P_{\infty}}{-\varphi \rho} \right)^{1/2} \left\{ 1-\left[ \frac{R(0)}{R(t)} \right]^\psi  \right\}^{1/2}
      \end{dcases}
       \ \ .
    \end{split}
    \label{fractal_roc_size}
\end{equation}
Hence the void collapsing time can be found to be:
\begin{itemize}
    \item For $\mathscr{D}_\text{s} \geq (1+\sqrt{3})/2$:
    \begin{equation}
\begin{split}
&\tau_{\text{c}} = \left( \frac{\varphi \rho}{2P_{\infty}} \right)^{1/2}\int_0^{R(0)} dR' \left\{ \left[ \frac{R(0)}{R'} \right]^\psi - 1 \right\}^{-1/2}  
\\
& \ \ \ \ \ \ \ \ = \frac{\displaystyle \left(\frac{\varphi \pi}{2} \right)^{1/2} \Gamma \left( \frac12 + \frac1\psi \right)}{\displaystyle \psi \Gamma \left( 1 + \frac1\psi \right)} \left[ \left( \frac{ \rho}{P_{\infty}} \right)^{1/2} R(0) \right] \ \ .
\end{split}
\label{fractal_collapse_inviscid}
\end{equation}
\\
\item For $1 \leq \mathscr{D}_\text{s} < (1+\sqrt{3})/2$:
    \begin{equation}
\begin{split}
&\tau_{\text{c}} = \left( \frac{-\varphi \rho}{2P_{\infty}} \right)^{1/2}\int_0^{R(0)} dR' \left\{ 1 - \left[ \frac{R(0)}{R'} \right]^\psi \right\}^{-1/2}  
\\
& \ \ \ \ \ \ \ \ = \frac{\displaystyle \left(\frac{-\varphi \pi}{2} \right)^{1/2} \Gamma \left( \frac1{-\psi} \right)}{\displaystyle -\psi \Gamma \left( \frac12 +\frac1{-\psi} \right)} \left[ \left( \frac{ \rho}{P_{\infty}} \right)^{1/2} R(0) \right] \ \ .
\end{split}
\label{fractal_collapse_inviscid_2}
\end{equation}
\end{itemize}
These cases both reveal a linear-dependent on the initial size of the sphere vacuum void $\tau_{\text{c}} \propto R(0)$, and the result $\tau_{\text{c}}$ only depends on the surface fractality $\mathscr{D}_{\text{s}}$. 

\begin{figure*}[!htbp]
\includegraphics[width=0.7 \textwidth]{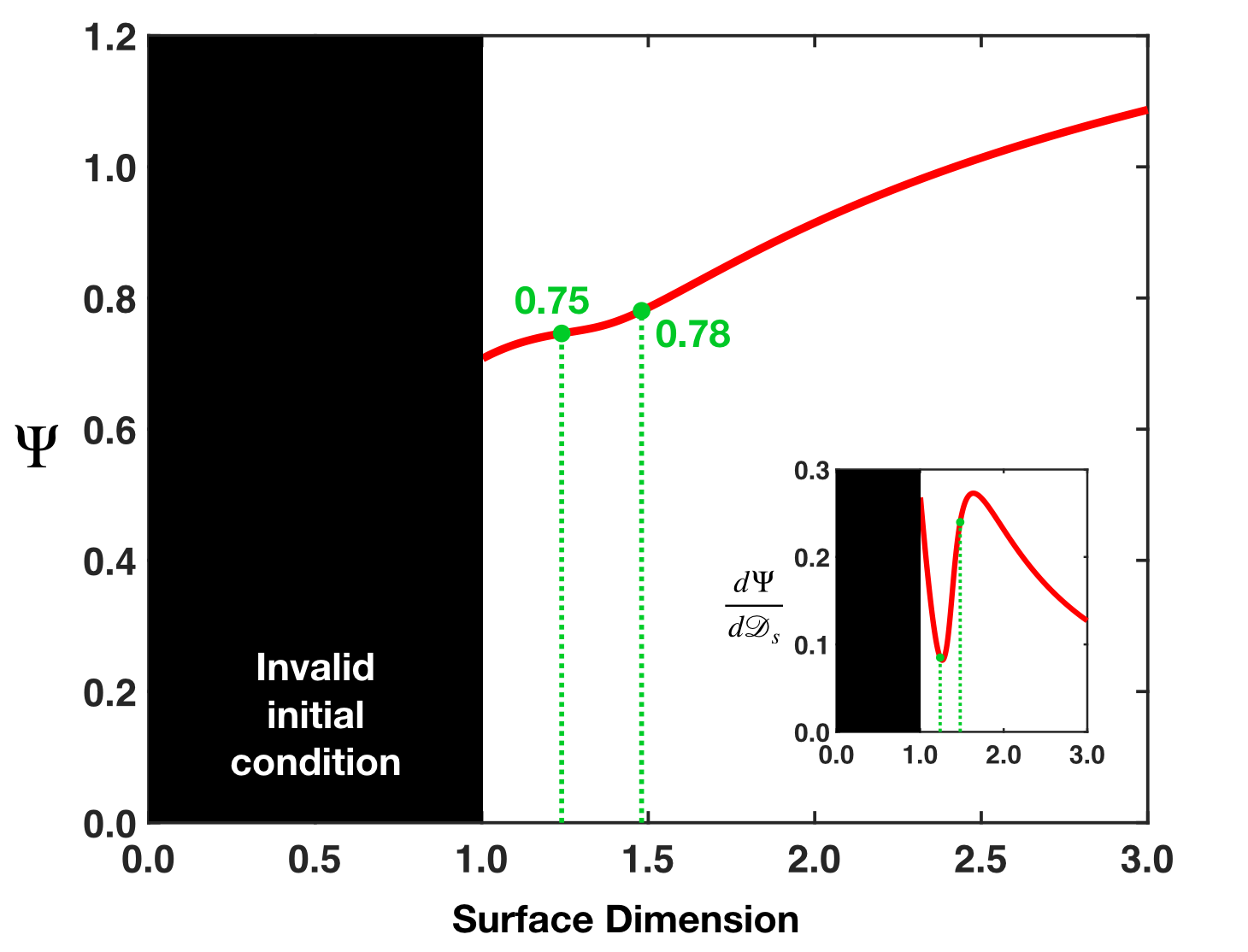}
\caption{The dimensionless coefficient $\Psi(\mathscr{D}_\text{s})$ (red line) of the total collapsing time $\tau_{\text{c}}$ as a function of the the surface dimension $\mathscr{D}_\text{s}$, $\displaystyle \tau_{\text{m}} = \Psi(\mathscr{D}_\text{s}) \left[ \left( \frac{ \rho}{P_{\infty}} \right)^{1/2} R(0) \right]$ as in Eq. \eqref{Psi}. The green lines locate the fractal dimensionalities of natural soils \cite{gimenez1997fractal}. There is a very non-trivial and quite subtle feature right inside this range, which can be seen clearest when considering the derivative $d\Psi/d\mathscr{D}_\text{s}$.}
\label{figS01}
\end{figure*}

Define the dimensionless coefficient in Eq. \eqref{fractal_collapse_inviscid} to be:
\begin{equation}
\begin{dcases}
     \mathscr{D}_\text{s} \geq \frac{1+\sqrt{3}}{2} \ &: \ \Psi (\mathscr{D}_\text{s}) = \frac{\displaystyle \left[\frac{\varphi(\mathscr{D}_\text{s}) \pi}{2} \right]^{1/2} \Gamma \left[ \frac12 + \frac1{\psi(\mathscr{D}_\text{s})} \right]}{\displaystyle \psi(\mathscr{D}_\text{s}) \Gamma \left[ 1 + \frac1{\psi(\mathscr{D}_\text{s})} \right]}
      \\[10pt]
      1 \leq \mathscr{D}_\text{s} < \frac{1+\sqrt{3}}{2} \ &: \ \Psi (\mathscr{D}_\text{s}) = \frac{\displaystyle \left[\frac{-\varphi(\mathscr{D}_\text{s}) \pi}{2} \right]^{1/2} \Gamma \left[  \frac1{-\psi(\mathscr{D}_\text{s})} \right]}{\displaystyle -\psi(\mathscr{D}_\text{s}) \Gamma \left[ \frac12 + \frac1{-\psi(\mathscr{D}_\text{s})} \right]}
      \end{dcases}
\ \ ,
\label{Psi}
\end{equation}
where the coefficients $\varphi(\mathscr{D}_\text{s})$, $\psi(\mathscr{D}_\text{s})$ are as defined in Eq. \eqref{coeffs}. We numerically investigate this in Fig. \ref{figS01}. {\color{black} For verification}, $\Psi(2)=0.91$, which agrees with Eq. \eqref{collapsing_time_3D}.

\subsection{Stokes Flow \label{supp_fractal_stokes}}

Consider the extreme viscous limit $\mu \rightarrow \infty$, Eq. \eqref{fractal_void_eom} becomes:
\begin{equation}
\frac{\displaystyle 2 (\mathscr{D}_\text{v} - \alpha_r) \Gamma\left(\frac{\alpha_r}{2}\right) \mu}{\displaystyle \pi^{\alpha_r/2} R^{\alpha_r} } \frac{d}{dt} R + \frac1\rho P_{\infty} \approx 0 \ \ .
\end{equation}
The rate of bubble size change reaches critical velocity $dR(t)/dt= \text{const}$ quickly when we work with this approximation. We then rearrange the differentials to obtain:
\begin{equation}
dt = \frac{\displaystyle 2 (\mathscr{D}_\text{v} - \alpha_r) \Gamma\left(\frac{\alpha_r}{2}\right)}{\displaystyle \pi^{\alpha_r/2} } \frac{\mu \rho}{P_\infty} \frac{dR}{R^{\alpha_r}} \ \ ,
\end{equation}
and the void collapsing time can be found by doing an integration:
\begin{equation}
\tau_{\text{c}} = \frac{\displaystyle 2 (\mathscr{D}_\text{v} - \alpha_r) \Gamma\left(\frac{\alpha_r}{2}\right)}{\displaystyle \pi^{\alpha_r/2} } \frac{\mu \rho}{P_\infty} \int_0^{R(0)} \frac{dR'}{{R'}^{\alpha_r}} = \frac{\displaystyle 2 (\mathscr{D}_\text{v} - \alpha_r) \Gamma\left(\frac{\alpha_r}{2}\right)}{\displaystyle \pi^{\alpha_r/2} \left( 1-\alpha_r \right) } \left[ \frac{\mu \rho}{P_\infty} R^{1-\alpha_r}(0) \right] \ \ ,
\label{fractal_collapse_viscous}
\end{equation}
which reveals a power-law dependency, on the initial cavity size with the radial fractal dimensionality, as $\tau_{\text{c}} \propto R^{1-\alpha_r}(0)$.

\subsection{Fig. \ref{fig03}A in ``Slow-Mo'' \label{slowmo_app}}

\begin{figure*}[!htbp]
\includegraphics[width=\textwidth]{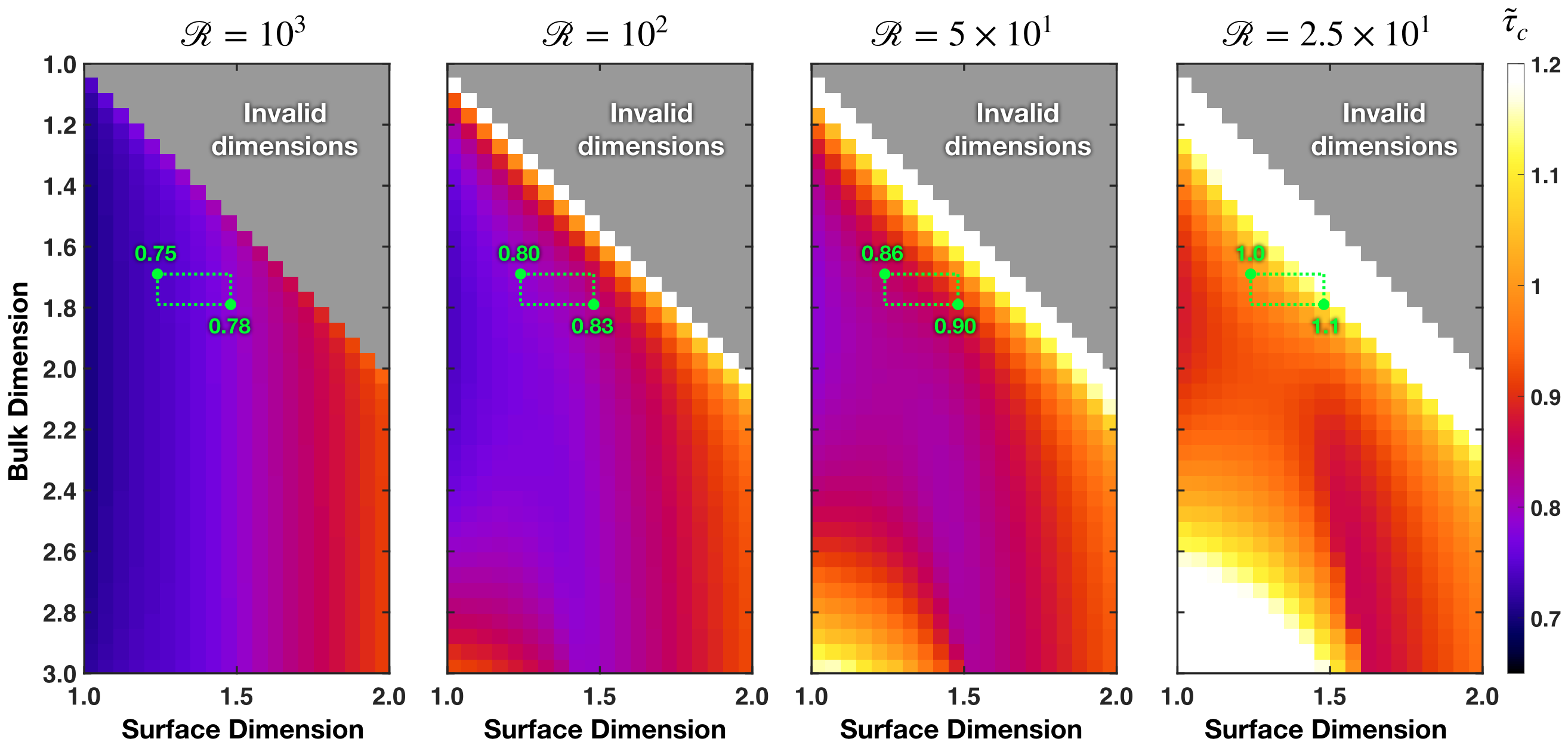}
\caption{\textbf{The collapsing time of vacuum void as we slowly vary the effective Reynolds number.} A more gradual progression of how the collapsing time surface $\tilde{\tau}_\text{c} \left( \mathscr{D}_\text{v},\mathscr{D}_\text{s}\right)$ changes as the effective Reynolds number $\mathscr{R}$ decreases. The green lines specify the dimensionalities of natural soils \cite{gimenez1997fractal} and the green values indicate the corresponding $\tilde{\tau}_{\text{c}}$.}
\label{figS02}
\end{figure*}

In Fig. \ref{figS02} we show how the surface $\tilde{\tau}_\text{c} \left( \mathscr{D}_\text{v},\mathscr{D}_\text{s}\right)$ changes with $\mathscr{R}$, more gradual compared to what have been shown in Fig. \ref{fig03}A.

\bibliography{main}
\bibliographystyle{unsrt}

\end{document}